\documentclass{aa}  

\usepackage{graphicx}
\usepackage{txfonts}
\begin{document}

   \title{AGN feedback in action in the molecular gas ring of the Seyfert galaxy NGC~7172}

   \titlerunning{AGN feedback in action in the molecular gas ring of
     the Seyfert galaxy NGC~7172}
   
   \author{A. Alonso Herrero
          \inst{1}
         \and
         S. Garc\'{\i}a-Burillo\inst{2}
         \and
         M. Pereira-Santaella\inst{3}
         \and
         T. Shimizu\inst{4}
         \and
         F. Combes\inst{5}
         \and
         E. K. S. Hicks\inst{6}
         \and
         R. Davies\inst{4}
         \and
         C. Ramos Almeida\inst{7, 8}
         \and
         I. Garc\'{\i}a-Bernete\inst{9}
         \and
          S. F. H\"onig\inst{10}
         \and
          N. A. Levenson\inst{11}
         \and
          C. Packham\inst{12, 13}
         \and
         E. Bellocchi\inst{14, 15}
         \and
         L. K. Hunt\inst{16}
         \and
         M. Imanishi\inst{13, 17}
         \and
          C. Ricci\inst{18, 19}
         \and
         P. Roche\inst{9}
       }

       \institute{Centro de Astrobiolog\'{\i}a (CAB), CSIC-INTA,
         Camino Bajo del Castillo s/n, E-28692 Villanueva de la Ca\~nada, Madrid,
     Spain\\
     \email{aalonso@cab.inta-csic.es}\label{inst1}
   \and
      Observatorio de Madrid, OAN-IGN, Alfonso XII, 3, E-28014 Madrid,
      Spain\label{inst2}
      \and
      Instituto de F\'{\i}sica Fundamental, CSIC, Calle Serrano 123, 28006 Madrid, Spain\label{inst3}
      \and
       Max-Planck-Institut f\"ur Extraterrestrische Physik, Postfach 1312,
       85741 Garching, Germany\label{inst4}
       \and
       LERMA, Observatoire de Paris, Coll\`ege de France, PSL University,
       CNRS, Sorbonne University, Paris, France\label{inst5}
       \and
       Department of Physics \& Astronomy, University of Alaska Anchorage,
       Anchorage, AK 99508-4664, USA\label{inst6}
       \and
        Instituto de Astrof\'{\i}sica de Canarias, Calle V\'{\i}a L\'actea, s/n, 38205 La
         Laguna, Tenerife, Spain\label{inst7}
         \and
         Departamento de Astrof\'{\i}sica, Universidad de La Laguna, 38206 La
         Laguna, Tenerife, Spain\label{inst8}
           \and
         Department of Physics, University of Oxford, Oxford OX1 3RH,
         UK\label{inst9}
       \and
    School of Physics \& Astronomy, University of Southampton,
    Southampton SO17 1BJ, Hampshire, UK\label{inst10} 
\and
Space Telescope Science Institute, 3700 San Martin Drive, Baltimore, MD 21218,
USA\label{inst11}
\and
The University of Texas at San Antonio, One UTSA Circle, San
Antonio, TX 78249, USA\label{inst12}
\and
National Astronomical Observatory of Japan, National Institutes of
Natural Sciences (NINS), 2-21-1 Osawa, Mitaka, Tokyo 181-8588,
Japan\label{inst13}
\and
Departamento de F\'{\i}sica de la Tierra y Astrof\'{\i}sica, Fac. de
CC F\'{\i}sicas, Universidad Complutense de Madrid, 28040 Madrid,
Spain\label{inst14}
\and
Instituto de F\'{\i}sica de Part\'{\i}culas y del Cosmos IPARCOS,
Fac. CC F\'{\i}sicas, Universidad Complutense de Madrid, 28040 Madrid,
Spain\label{inst15}
\and
INAF – Osservatorio Astrofisico di Arcetri, Largo Enrico Fermi 5,
50125 Firenze, Italy\label{inst16}
\and
Department of Astronomy, School of Science, Graduate
University for Advanced Studies (SOKENDAI), Mitaka, Tokyo 181-8588,
Japan\label{inst17}
\and
N\'ucleo de Astronom\'ia de la Facultad de Ingenier\'ia, Universidad Diego Portales, Av. Ej\'ercito Libertador 441, Santiago, Chile\label{inst18}
\and
Kavli Institute for Astronomy and Astrophysics, Peking University, Beijing 100871, China\label{inst19}
}

   \date{Received 2023; accepted 2023}

  \abstract
{We present new ALMA observations of the CO(3-2) transition
  and associated $854\,\mu$m continuum at $0.06-0.3\arcsec$ resolution, together with new VLT/SINFONI
observations of  NGC~7172. This is a luminous (bolometric luminosity
of $\simeq 10^{44}\,{\rm erg\,s}^{-1}$) Seyfert galaxy
that belong to the Galaxy Activity, Torus, and Outflow Survey (GATOS). The ALMA CO(3-2)
observations reveal the presence of a highly inclined cold molecular
gas ring with an
approximate radius of $3-4\arcsec \simeq 540-720\,$ pc, which is likely
associated with an inner Lindblad resonance of a putative stellar bar.  There are noncircular motions in
the  VLT/SINFONI [Si\,{\sc vi}]$\lambda$$1.96\,\mu$m and H$_2$ at
$2.12\,\mu$m, and ALMA 
CO(3-2) velocity fields. After subtracting the stellar velocity field,
we detected  [Si\,{\sc
  vi}] blueshifted velocities of a few hundred
 ${\rm km\,s}^{-1}$ to the south of the active galactic nucleus (AGN) position. They trace
 outflowing ionized gas outside the plane of the galaxy and out to
 projected distances of $\simeq 200\,$pc. The CO(3-2)
 position-velocity diagram along the kinematic minor 
  axis displays noncircular motions with observed
  velocities of up to $\sim 150\,{\rm km\,s} ^{-1}$. Assuming that
  these are taking place in the disk of the galaxy, the observed
  velocity signs imply that the molecular gas ring is not only rotating
  but also outflowing. We derived an  integrated cold molecular gas mass outflow
  rate of $\sim 40\,M_\odot\,{\rm yr}^{-1}$ for the ring. 
Using the ALMA $854\,\mu$m extended emission map, we resolved a 
32\,pc radius torus with a gas mass of
 $8\times 10^5\,M_\odot$. These torus properties
 are similar to other Seyfert galaxies in the GATOS sample. We 
 measured a decreased cold molecular gas concentration in the nuclear-torus region
 relative to the circumnuclear region when compared to other less luminous
 Seyfert galaxies. We conclude that  the effects of AGN
feedback in NGC~7172, which are likely caused by the AGN wind and/or the moderate luminosity radio
jet, are seen as a 
large-scale outflowing molecular gas ring and accompanying 
redistribution of molecular gas in the nuclear regions.}


   \maketitle
%

\section{Introduction}\label{sec:introduction}

The formation and evolution of galaxies is one of the key challenges
of modern astrophysics. State-of-the-art cosmological simulations
provide the theoretical
benchmark to understand the growth of galaxies and their supermassive
black holes, and in particular, the effects of feedback \citep[see for
instance,][]{Sijacki2015, Dubois2016, Dave2019, AnglesAlcazar2021}.  
With the new computational advances, these simulations are able to
track the gas cycle on all relevant physical scales, from the
circumgalactic medium  all the way down to the sphere of
influence of supermassive black holes. One important result from these
simulations is that feedback is needed to regulate the buildup of the
stellar mass in massive galaxies. In these types of galaxies, feedback
is believed to be 
driven by radiation pressure from an active galactic nucleus (AGN) and
mechanical energy from  radio jets \citep[see e.g.,][for a
review]{Fabian2012}. 

Energetic multiphase (ionized, neutral, warm,
and cold molecular gas) outflows have been routinely discovered in
powerful AGNs and (ultra)luminous infrared 
galaxies. In strongly star-forming galaxies, the
molecular outflow rates are comparable to, or higher by a factor of
few, than the ongoing star-formation activity \citep{Cicone2014}. When an AGN is
present, the outflow rates can be strongly boosted, with more luminous AGNs
driving more energetic outflows \citep[see, 
e.g.,][]{Fiore2017, Fluetsch2019}. Still, at a given AGN luminosity, the outflow
rates of the ionized gas span a range of nearly two orders of
magnitude in nearby Seyfert galaxies
\citep[see][and references therein]{Davies2020}. Similar results are
found for  the molecular
phase of outflows in AGNs and ultraluminous infrared galaxies (ULIRGs), where
the AGN luminosity is found not to be 
the only factor driving massive outflows \citep[see e.g.,][respectively]{RamosAlmeida2022,
  Lamperti2022}. Other factors, such as jet and/or wind inclination relative 
to the galaxy disks, or gas concentration can also play a significant
role.  

In nearby Seyfert galaxies where the molecular and ionized gas phases
were studied together, there is evidence for positive and/or negative
feedback effects on scales of out to a few kiloparsecs caused by AGN
winds, sometimes combined with low-power radio jets
\citep{GarciaBurillo2014, Cresci2015, Shimizu2019, Venturi2018,  
  Venturi2021, GarciaBernete2021, Gao2021}. At Seyfert-like
luminosities, the molecular phase    
has a larger contribution to the total mass outflow rate and kinetic power than the ionized
phase \citep{Fiore2017, Fluetsch2019}. The superb angular resolution attained with
(sub)millimeter interferometry, including the Atacama Large
Millimeter/submillimeter Array (ALMA), allows for AGN-driven cold
molecular gas outflows to be probed from circumnuclear scales of a few
hundred parsecs all the way to the molecular dusty torus on tens of
parsecs. The derived molecular mass outflow rates in Seyfert galaxies range from $\sim
1\,M_\odot$ yr$^{-1}$ to a few tens of $M_\odot$ yr$^{-1}$, and the
velocities are up to  hundreds of km s$^{-1}$ \citep{Combes2013, GarciaBurillo2014, Morganti2015,
  Zschaechner2016, AlonsoHerrero2018,
  AlonsoHerrero2019, Audibert2019, DominguezFernandez2020}.

 The Galactic Activity, Torus, and Outflow Survey
(GATOS) aims to understand the obscuring material (torus) and the nuclear gas cycle
(inflows and outflows) in the dusty regions immediately surrounding the
active nucleus of nearby AGNs.  To do so, we selected local Seyfert galaxies 
from the 70  month {\it Swift}/BAT All-sky Hard X-ray Survey
\citep{Baumgartner2013}. \cite{GarciaBurillo2021}
obtained ALMA band 7 observations of ten GATOS Seyfert galaxies in the southern hemisphere and
with distances in the 10-30\,Mpc range. Combining these Seyfert galaxies
with a sample of low-luminosity AGNs \citep[from][]{Combes2019}, 
they found that galaxies with
higher AGN luminosities and/or Eddington ratios are more efficient at
clearing molecular gas from the nuclear (torus) regions. This was
interpreted as the imprint on the nuclear molecular gas distribution left by
the presence of AGN winds, that is, AGN-driven feedback. In a few
GATOS Seyfert galaxies, this is confirmed by the presence of
(circum)nuclear molecular outflows \citep{AlonsoHerrero2018, AlonsoHerrero2019,
  GarciaBurillo2019} and polar dust emission
\citep{AlonsoHerrero2021} on these scales.

   \begin{figure*}
    \vspace{-0.75cm}
     \hspace{1.5cm}
     \includegraphics[width=16cm]{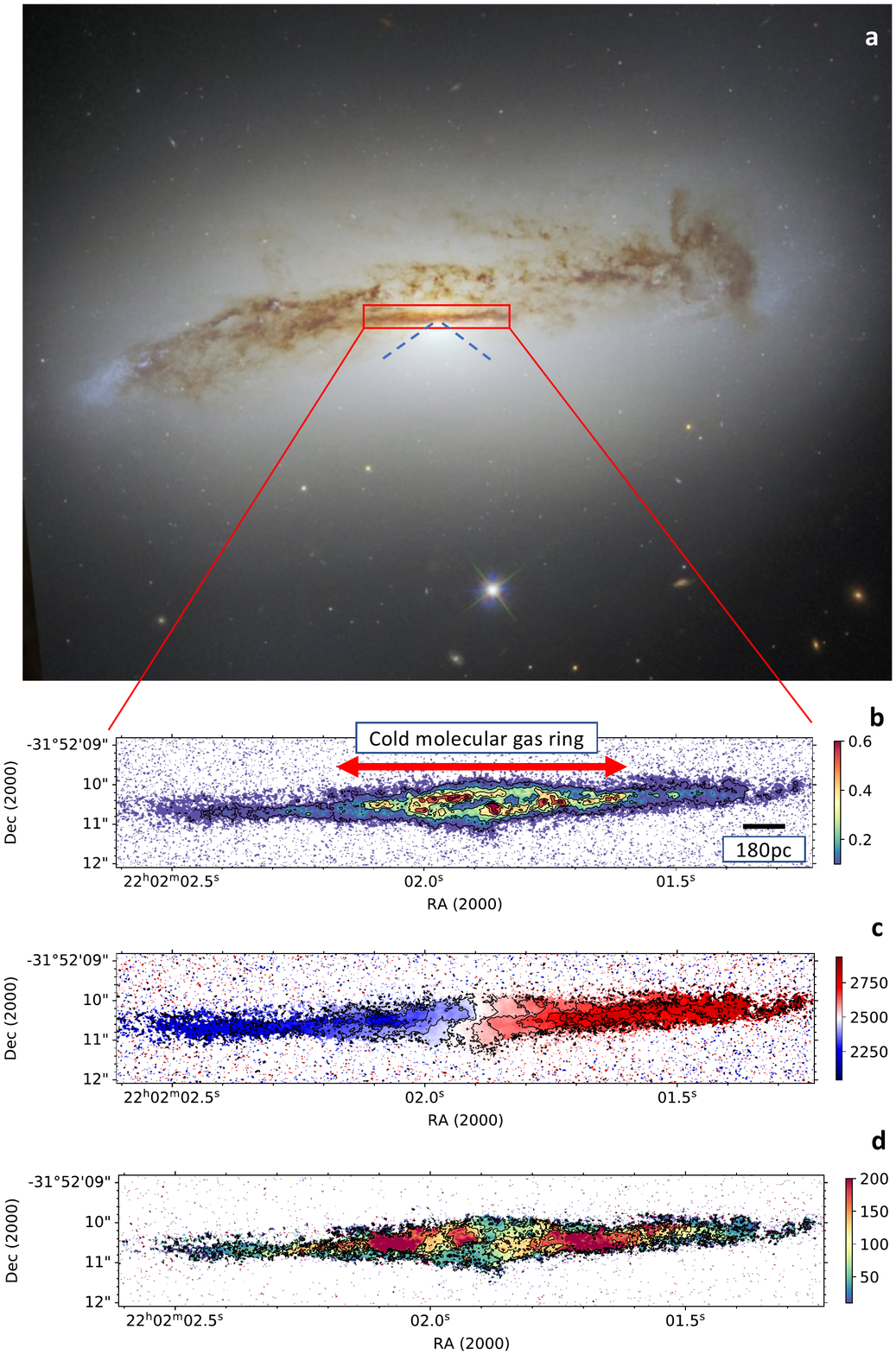}
     \vspace{-0.5cm}
     \caption{Seyfert galaxy NGC~7172. From top to bottom. {\it
       Panel a}: HST optical to near-IR
     composite RGB image of NGC~7172 retrieved from ESASky4.3.0. It
     combines observations from the Advanced Camera for Surveys (ACS)
     at 6060 and 8140\,\AA,
     and the Wide Field Camera 3 (WFC3) at $1.6\,\mu$m. The orientation
     of the image is north up, east to the left. The FoV
     is $\simeq$2\arcmin $\times$ 2\arcmin. The rectangle shows
    approximately  the ALMA FoV, while the dashed lines represent the
    opening angle 
    of the southern (brighter) side of the optical ionization cone. 
     Image credit:
     ESA/Hubble \& NASA, D. J. Rosario, A. Barth. {\it Panel b:} ALMA
     CO(3-2) intensity map (moment 0) produced with a 3$\sigma$
     clipping using the merged configuration observations. We trimmed
     the original FoV to
     approximately $17.6\arcsec \times 3.3\arcsec$. Both
     the image and the contours are shown
     on a linear scale. The color bar shows the intensity scale in Jy km
     s$^{-1}$ per beam units. The first contour is at 0.05\,Jy km
     s$^{-1}$ per beam. The ALMA beam (not shown) is $0.08\arcsec \times
     0.07\arcsec$ ($14\,{\rm pc} \times 13\,{\rm pc}$) at PA$_{\rm
       beam}=90.4^\circ$. {\it Panel c:} ALMA
     CO(3-2) mean velocity map (moment 1) produced with a 3$\sigma$
     clipping using the merged configuration observations. The
     contours are the isovelocities. {\it Panel d:} ALMA
     CO(3-2) mean velocity dispersion map (moment 2) produced with a 3$\sigma$
     clipping using the merged configuration observations. For the
     last two panels the scale bar units are ${\rm km\,s}^{-1}$.}
              \label{fig:HSTALMAlarge}
    \end{figure*}

    Up until now, we investigated the ALMA properties of
    intermediate luminosity Seyfert galaxies with $L_{\rm
  AGN}(2-10\,{\rm keV})=10^{41.4-43.5}\,{\rm erg\,s}^{-1}$ and
distances of less than 28\,Mpc in the
GATOS sample \citep{AlonsoHerrero2018, AlonsoHerrero2019, GarciaBurillo2021}.
NGC~7172 is a nearly edge-on Sa spiral galaxy in the GATOS sample at
a redshift-independent distance  of 37\,Mpc \citep{Davies2015} with an
AGN bolometric luminosity of $\sim 1.3 \times
10^{44}\,$erg s$^{-1}$ \citep{Davies2015}.  It
belongs to the Hickson compact group HCG90, which also includes,
among other galaxies, NGC~7173, NGC~7174, and NGC~7176.
NGC~7172 is optically classified as a Seyfert 2
\citep{VeronCetty2006}, although \cite{Smajic2012} found evidence for the presence of
broad hydrogen recombination lines in the near-infrared (near-IR).
The nuclear region is crossed by  dust lane that obscures the galaxy nucleus in the optical
but not at longer wavelengths \citep{Smajic2012}. There is also a broader
dust lane extending east-west across the galaxy that appears to be warped at the edges
\citep{Sharples1984}.    Figure~\ref{fig:HSTALMAlarge} (top panel) shows a
red-green-blue (RGB) 
optical to near-IR {\it Hubble}
Space Telescope (HST) image of NGC~7172 retrieved from ESASky
\citep{Baines2017, Giordano2018}, where this morphology is clearly
seen.  The  distribution of the prominent dust lanes to the north
of the nucleus indicates this is the near side of the galaxy. This is 
also in agreement with the extinction distribution in the
(circum)nuclear regions of the galaxy \citep{Smajic2012}.

\cite{Thomas2017} detected a prominent two-sided ionization cone in NGC~7172
using optical integral field unit (IFU) spectroscopy. It has a projected opening
angle of approximately $120^\circ$. The southern part of the cone is 
brighter, which is consistent with the position of the main dust
lane and with north being the near side of the galaxy (see
Fig.~\ref{fig:HSTALMAlarge}, top panel). The size on this 
side of the cone is approximately $5\arcsec$,
which is equivalent to a projected physical size of 900\,pc. Using spectroscopic observations with
X-shooter on the Very Large Telescope (VLT),  \cite{Davies2020} derived a velocity for the ionized gas
outflow of $\sim 400\,{\rm km\,s}^{-1}$ and an
outflow rate of 0.005\,M$_\odot$ yr$^{-1}$.
In this work, we present
new band 7 ALMA observations of the  CO(3-2) transition and the associated
$854\,\mu$m continuum  of the GATOS  galaxy
NGC~7172. The ALMA observations were taken with 
an angular resolutions of approximately 0.06\arcsec \, and
0.3\arcsec, which correspond to 11\,pc and 54\,pc, respectively for
the assumed distance. This allows us to resolve the morphology and
kinematics of the cold molecular gas of  both the nuclear and
circumnuclear regions of this galaxy.
This paper is organized as follows. Section~\ref{sec:observations}
presents the ALMA band 7 CO($J=3-2$) and associated continuum observations as
well as ancillary new VLT/SINFONI observations. In Sect.~\ref{sec:morphology} we
discuss the overall ALMA and SINFONI morphology and kinematics, while in
Sect.~\ref{sec:kinematics} we model the ALMA CO(3-2) kinematics. In
Sect.~\ref{sec:torus}, we derive the
properties of the  torus of
NGC~7172. Finally, in Sects.~\ref{sec:discussion} and
\ref{sec:conclusions} we discuss the results of 
this work and present our conclusions, respectively.

\begin{figure*}
\hspace{-0.2cm}
     \includegraphics[width=9.cm]{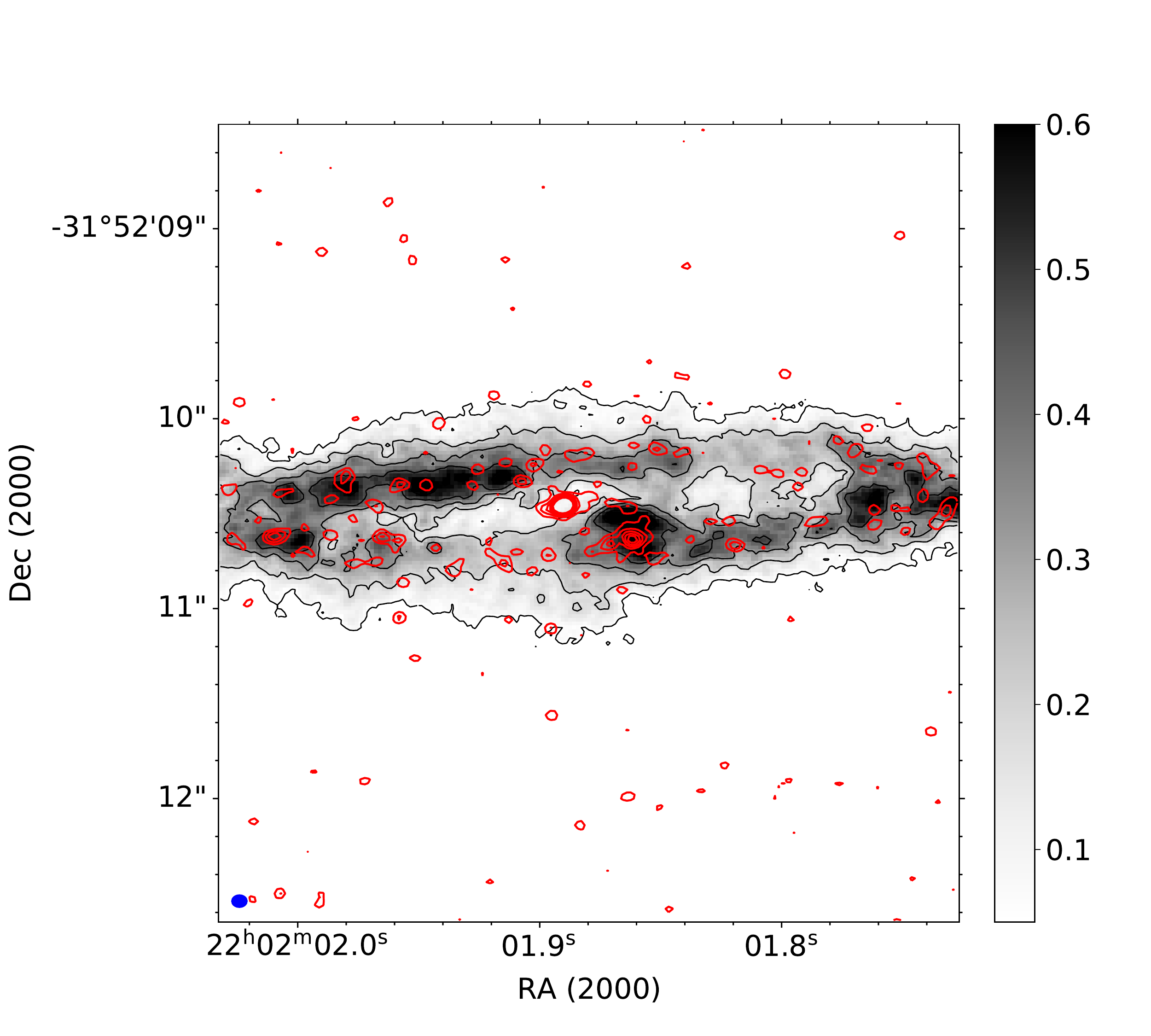}
\hspace{0.5cm}
\includegraphics[width=9cm]{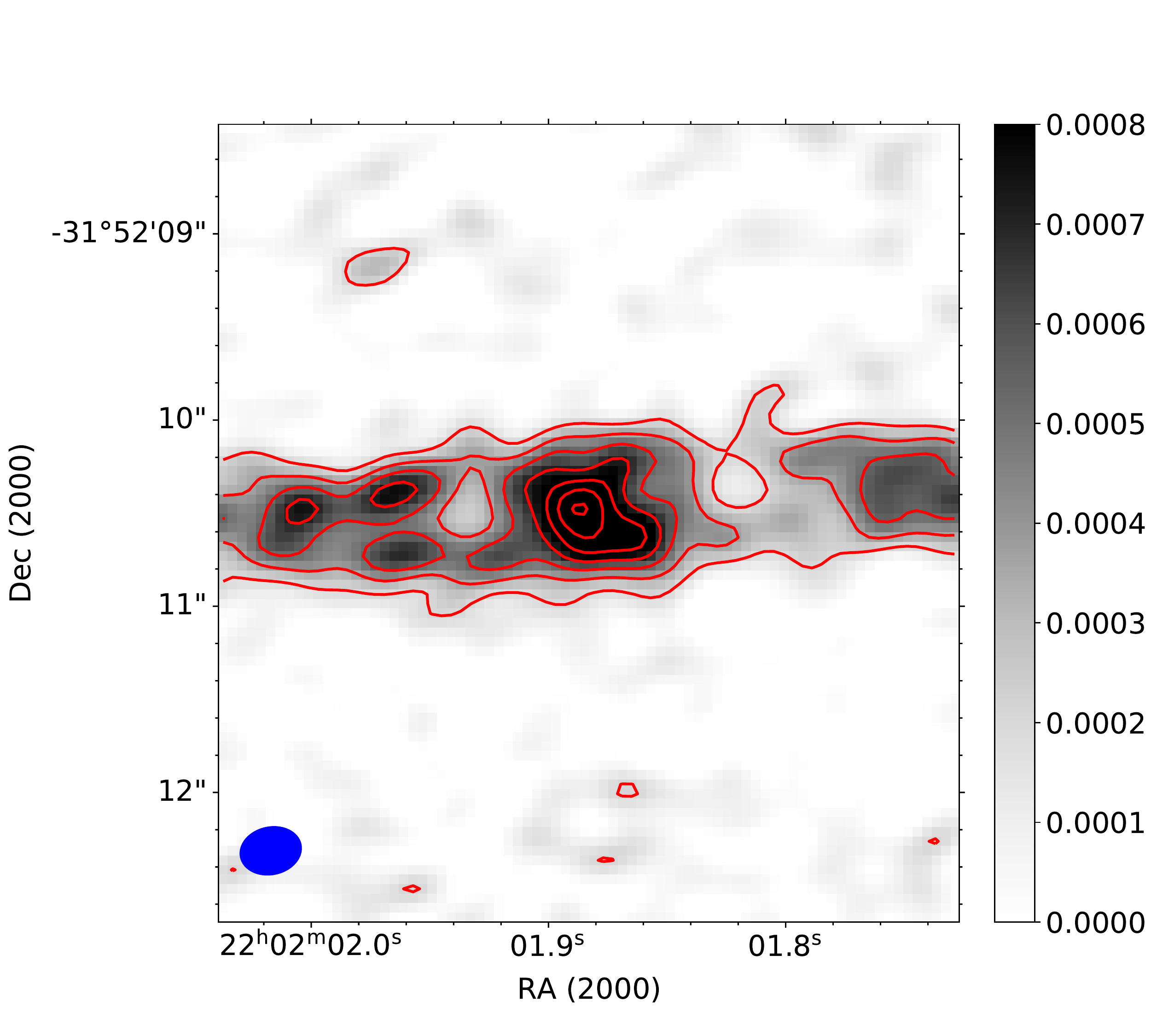}
\vspace{-0.1cm}
\caption{ALMA band 7 observations.
  {\it Left panel}: In gray scale and contours is a zoom-in on the ALMA
       CO(3-2) intensity map of Fig.~\ref{fig:HSTALMAlarge} (second
       panel from the top)
       showing
       the inner part of the cold molecular gas ring. The FoV
       matches that of the SINFONI observations (see
       Fig.~\ref{fig:ALMASINFONI}). The gray bar shows the CO(3-2) intensity scale in Jy km
     s$^{-1}$ beam$^{-1}$. The
       red contours are the band 7 continuum map at $854\,\mu$m from
       the merged configuration, shown on a
       linear scale.
       The first contour is at three times the measured continuum rms of
       $17\,\mu$Jy. The continuum beam ($0.08\arcsec \times
     0.06\arcsec$ at PA$_{\rm beam}=90.6^\circ$) is the blue filled ellipse in
     the bottom-left corner.
   {\it Right panel}: The gray scale map and the red contours are the compact
   configuration $854\,\mu$m continuum map, with the same FoV as in
   the left panel. The contours are on a linear scale. The continuum beam ($0.34\arcsec \times
     0.27\arcsec$ at PA$_{\rm beam}=-78^\circ$) is the blue filled ellipse in
     the bottom-left corner.}
              \label{fig:RINGALMACOcont}%
    \end{figure*}

\begin{table}
\caption{ALMA band 7 observations.}\label{tab:ALMAobservations}
\centering
\begin{tabular}{lcccc}
\hline\hline
Configuration & Continuum/&beam& PA$_{\rm beam}$ & rms\\
& Line& ($\arcsec \times \arcsec$) & ($^\circ$)\\
  \hline
Compact & $854\,\mu$m   &$0.34\times 0.27$ & $-78$ & 97\\
              & CO(3-2) & $0.33 \times 0.25$ & $-75$ &  2\\

  \smallskip
Extended & $854\,\mu$m & $0.07 \times 0.06$& $-88$ & 30\\

  \smallskip
 Merged & $854\,\mu$m    &$0.08\times 0.06$ & 91 & 17\\
              & CO(3-2) & $0.08 \times 0.07$& 90 & 0.5\\
  
  \hline
\end{tabular}
\tablefoot{Details of the beam sizes and PA as well as the rms of each of
  the configurations used in this wok. The rms units are $\mu$Jy per
  beam for the $854\,\mu$m continuum and mJy
 per channel and per beam for the CO(3-2) observations.}
\end{table}

    \section{Observations and data reduction}\label{sec:observations}
       \subsection{ALMA band 7}\label{subsec:alma}

We observed the emission of the CO(3-2) and HCO$^+$($J=4-3$) lines and their underlying continuum emission in the circumnuclear disk (CND) of NGC~7172 using  the band~7 ALMA receiver and a single pointing (project-ID:\#2019.1.00618.S; PI: A. Alonso-Herrero).  
 Although the phase tracking center of the galaxy was  assumed to be
 at right ascension and declination (RA$_{2000}$,
 Dec$_{2000}$)~=~(22$^{\rm h}$02$^{\rm
   m}$01.90$^{\rm s}$, --31$^{\rm o}$52'11.6''), the position of
 the AGN has nevertheless  been reassigned  through a comparison with
 the radio coordinates and a fit of the
 continuum emission, as detailed in Sects.~\ref{subsec:ALMAmorphology} and
 ~\ref{sec:torus}, respectively.  The ALMA   field of view (FoV) of
 17$\arcsec$ corresponds to the central 3.1~kpc region of the galaxy.
 We combined two sets of configurations of the ALMA array, namely, an extended
 (C43-7) and  
 a compact (C43-4) configuration. The aim was to  reach an angular
 resolution of $<0\farcs07=13$~pc while keeping the largest angular
 scale  recovered in our maps out to $\sim3\arcsec=540$~pc. This scale
 is enough to
 recover a sizeable fraction of the flux inside the FoV and image with
 high fidelity the torus and its connections with the CND region, which are relevant to this work.
Observations required the execution of two scheduling blocks for the C43-7
configuration set conducted in July 2021 and one  for the C43-4
conducted in October 2019, and the use of
$\sim43-44$ antennas of ALMA.

We placed four spectral windows  of 1.875~GHz-bandwidth, two in the lower side band (LSB) and two in the upper sideband (USB). This setup allowed us to observe the  CO($J=3-2$) line (345.796~GHz at rest) and the continuum emission (344.0~GHz at rest) in
the LSB bands, as well as  HCO$^+$(4-3) (356.734~GHz at rest) and
the continuum emission (358.1~GHz at rest) in the USB bands. We
calibrated the data making use of the ALMA 
reduction package  {\tt CASA} \citep{McMullin2007}\footnote{http//casa.nrao.edu/}.

In this work we make use of the line and continuum data sets from the compact, extended, and the merged configurations. The calibrated uv-tables of the merged configurations were exported to {\tt GILDAS\footnote{http://www.iram.fr/IRAMFR/GILDAS}}  to proceed
with the continuum subtraction and imaging procedures, as detailed
below. We first subtracted the continuum from each of the spectral
$(u,v)$ data sets using the {\tt GILDAS} task {\tt UV-BASELINE}. We
fit  a baseline to the  $(u,v)$ data sets through a polynomial of degree zero masking the line emission around each transition with a range of velocity widths $\sim900$~km~s$^{-1}$ and subsequently obtained continuum-free spectral line images for  CO(3-2) and
HCO$^+$(4-3). An inspection of the HCO$^+$ line data cube showed
no significant emission inside the ALMA FoV. The estimated upper
  limit flux for this transition is $0.12\,$Jy\,${\rm km\,s}^{-1}$, which provides a 
  CO(3-2) to HCO$^+$(4-3) ratio of $>1.4$. The lower upper limit
on the CO-to-HCO$^+$ ratio is 
  consistent with the values and upper limit measured in other GATOS
  Seyfert galaxies of similar Eddington ratios \citep{GarciaBurillo2021}.

In the rest of this work, we restrict our
analysis and discussion to the CO(3-2) line data.  We derived images
of the continuum emission by averaging in each of 
the two subbands centered around spectral lines those channels free of line emission using the  {\tt GILDAS} tasks {\tt UV-FILTER} and {\tt UV-CONT}, making use of the same velocity width masks employed by the {\tt UV-BASELINE} task. The four line-free
continuum uv-tables were combined using the task {\tt UV-MERGE}
to obtain a genuine continuum image of the galaxy at an average
frequency range $\sim351.1$~GHz or $854\,\mu$m (at rest).

For the merged data set,  we obtained an angular resolution $\simeq
0\farcs06-0\farcs08=11-14$~pc by changing in the {\tt GILDAS} task
{\tt UV-MAP} the robust parameter ($b$) to 1. 
The line data cubes were binned to a common frequency  resolution of
7.8~MHz (equivalent to $\sim$6.8~km~s$^{-1}$ in band~7). We estimated
that the flux accuracy is about 10--15$\%$, comparable to the level of
uncertainty of standard ALMA observations at the 
same range of frequencies. The point-source sensitivities in the line
data cubes were derived by selecting areas free from emission in all
channels. We summarize in Table~\ref{tab:ALMAobservations} all the
relevant parameters of the ALMA observations. We created CO(3-2)
moment 
maps from the merged configuration data set using the  {\tt GILDAS}
$\tt moment$ task with a 3$\sigma$ clipping. We show the 0th, 1st, and 2nd
moment maps in Fig.~\ref{fig:HSTALMAlarge},
while Fig.~\ref{fig:RINGALMACOcont} show the $854\,\mu$m continuum
maps of the central region observed with the compact and merged configurations.
Additionally, in Sect.~\ref{sec:kinematics} we discuss the moment maps
from the compact configuration. 

\subsection{VLT/SINFONI}\label{subsec:sinfoni}

In this section we present new near-IR IFU spectroscopic
observations obtained with the SINFONI instrument
\citep{Eisenhauer2003, Bonnet2004} at
the VLT,  as part of program 
093.B-0057(B) (PI: R. Davies).  NGC 7172 was observed during the
nights of 16 and 20 July
2014. The $H+K$ grating was used with laser guide star adaptive optics
to achieve a spectral resolution $R \simeq 1500$ over a $\simeq
3\arcsec \times 3\arcsec$ FoV. A standard
observing sequence of Object-Sky-Object was used with detector
integration times of 300 (12 exposures) and 50 seconds (6 exposures)
and small dithering between each Object exposure resulting in a total
exposure time of 65 minutes. 

The raw detector images were reduced with SPRED \citep{Abuter2006}, a
custom data reduction package for SINFONI developed at MPE. SPRED
performs all standard reduction steps needed to reconstruct NIR IFU
cubes. We further  used the MXCOR and SKYSUB routines
\citep{Davies2007} to improve OH sky emission subtraction and LAC3D, a 3D
version of LACOSMIC \citep{vanDokkum2001} to identify and remove bad
pixels and cosmic rays. We removed atmospheric telluric features and
flux calibrated the data based on observations of four B-type stars. A
final additional step we implemented was the correction for
differential atmospheric refraction which induces wavelength dependent
spatial offsets. The custom procedure for this is described in
\cite{Lin2018}.

\begin{figure*}

\vspace{-1cm}
\hspace{1cm}
      \includegraphics[width=17cm]{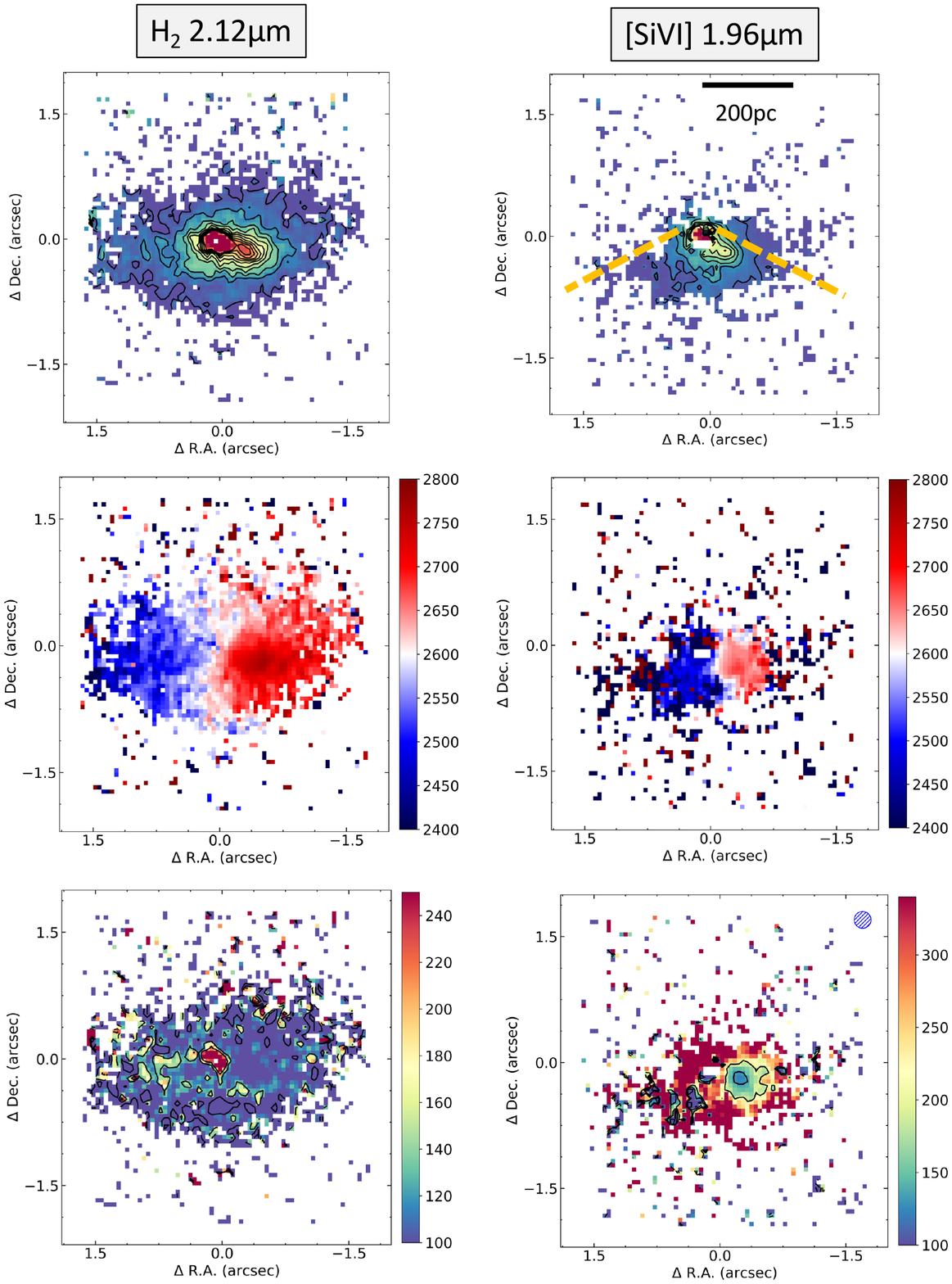}
\vspace{-0.75cm}
     \caption{VLT/SINFONI observations. From top to bottom,  integrated flux (in arbitrary
       units), velocity (in ${\rm km\,s}^{-1}$), and
       velocity dispersion (in ${\rm km\,s}^{-1}$) maps of the
       H$_2$ line at $2.12\,\mu$m ({\it left panels}) and [Si\,{\sc vi}]$\lambda
       1.96\,\mu$m line ({\it right panels}). In the top-right panel, the thick orange dashed lines
       indicate the approximate opening angle of the [O\,{\sc iii}]$\lambda$5007
       ionization cone \citep{Thomas2017}.
       We note that in these SINFONI maps, the fitted emission
       does not fully cover the $3.9\arcsec \times 4.2\arcsec$
       FoV. The colors and the contours are
       on a linear
       scale. The (0, 0) point
       in all the panels corresponds to 
        the peak of the SINFONI line maps. The hatched circle in the
        bottom-right panel shows the angular resolution (FWHM) of the
        observations.}
              \label{fig:SINFONI}%
    \end{figure*}

After the data reduction, the FoV of the SINFONI images is $3.9\arcsec \times 4.2\arcsec$
   with a pixel size of 0.05\arcsec.   We estimated an angular
   resolution of the observations of $\simeq 0.2\arcsec$ full width
   at half maximum (FWHM), as measured
   from the $K$-band continuum image (see below) of the galaxy. For this study, we focus on two
   emission lines, namely, the coronal line [Si\,{\sc vi}] at
     $1.96\,\mu$m to trace the ionized gas excited by the AGN
     \citep[see, e.g.,][]{MuellerSanchez2011, RamosAlmeida2017, Shimizu2019} and 
     H$_2$(1–0)S(1) $2.122\,\mu$m to trace the hot molecular gas
     emission. Before we fit these two emission lines in the
     SINFONI data   cube, we modeled and removed the continuum using the Penalized Pixel
  Fitting tool \citep[{\sc ppxf}, see ][]{Cappellari2004}. We modeled the emission
  lines using a single Gaussian to derive the flux, velocity, and
  velocity dispersion (corrected for instrumental resolution) maps
  shown in Fig.~\ref{fig:SINFONI}. 

 \begin{figure*}

\hspace{1cm}
      \includegraphics[width=7.5cm]{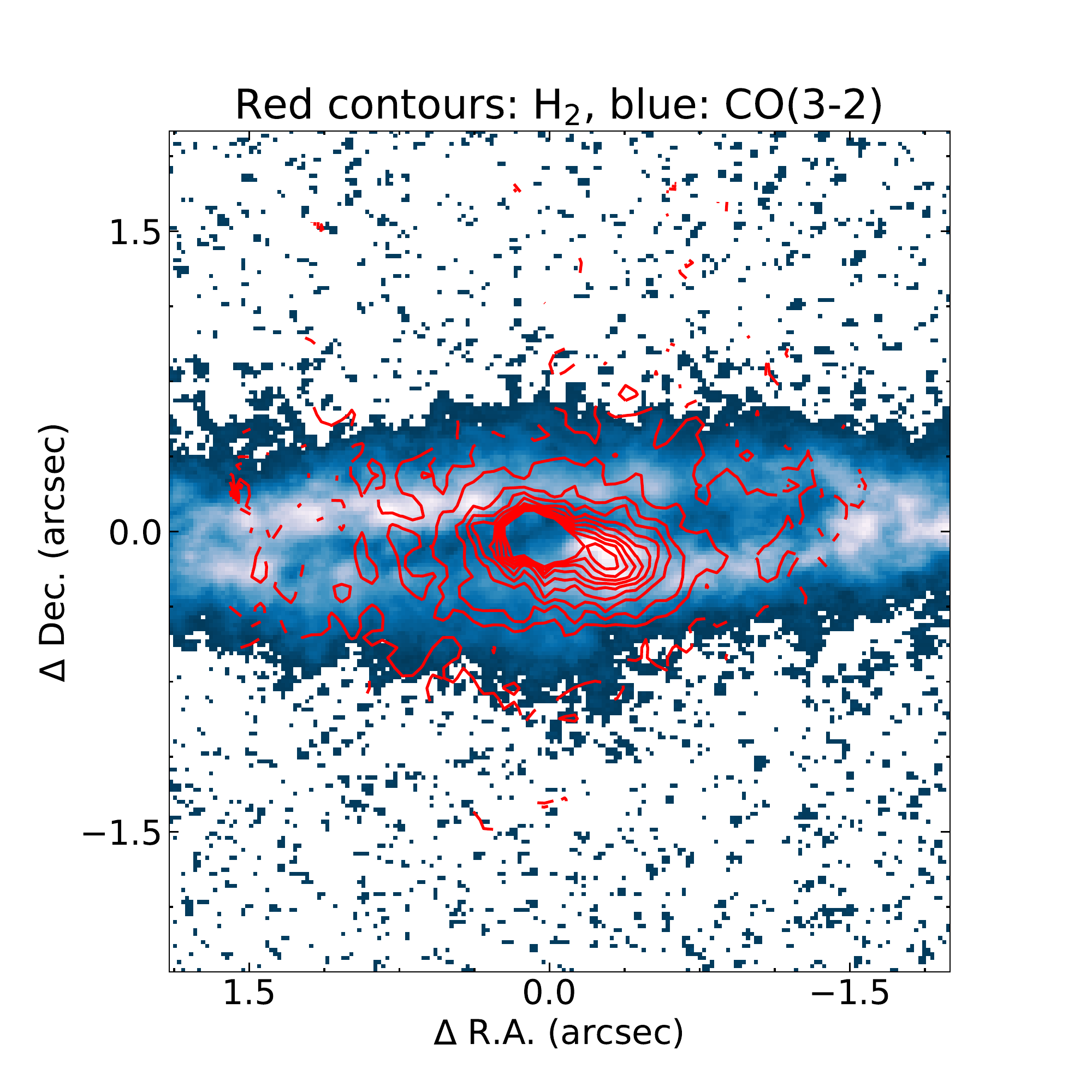}
      \includegraphics[width=7.5cm]{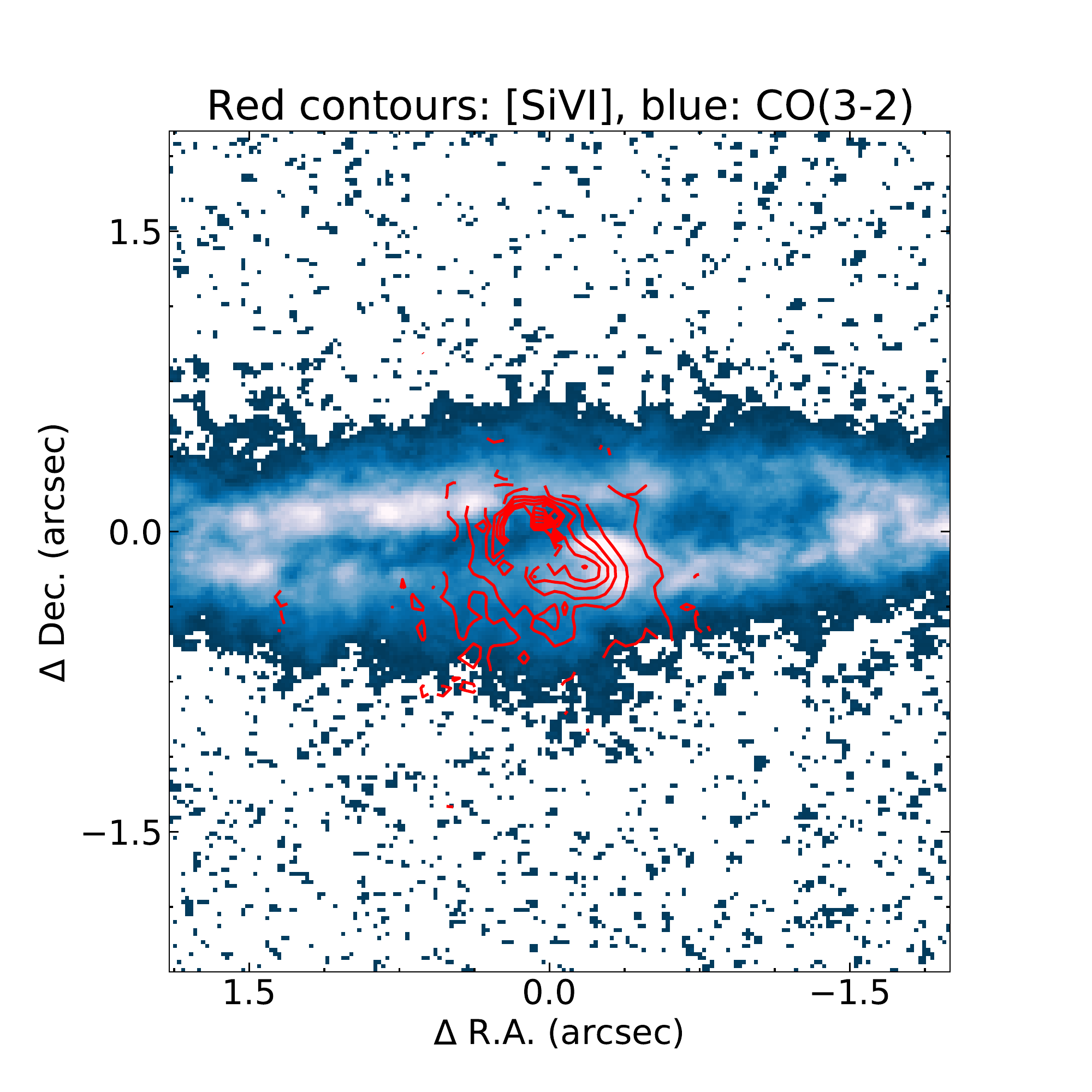}
     \caption{Comparison of the ALMA and VLT/SINFONI
         observations. The blue color scale image is a zoom-in on the ALMA
       CO(3-2) intensity map of Fig.~\ref{fig:HSTALMAlarge} with a FoV
       matching that of the VLT/SINFONI observations (see
       Fig.~\ref{fig:SINFONI}). The red contours are the 
     H$_2$ line at $2.12\,\mu$m ({\it left panel}) and [Si\,{\sc
       vi}]$\lambda 1.96\,\mu$m ({\it right panel}), as Fig.~\ref{fig:SINFONI}. The (0, 0) position
       in all the panels corresponds to 
        the peak of the SINFONI line maps as well as the ALMA
       $854\,\mu$m continuum.}
              \label{fig:ALMASINFONI}%
    \end{figure*}

 \begin{figure*}

\hspace{1cm}
      \includegraphics[width=7.5cm]{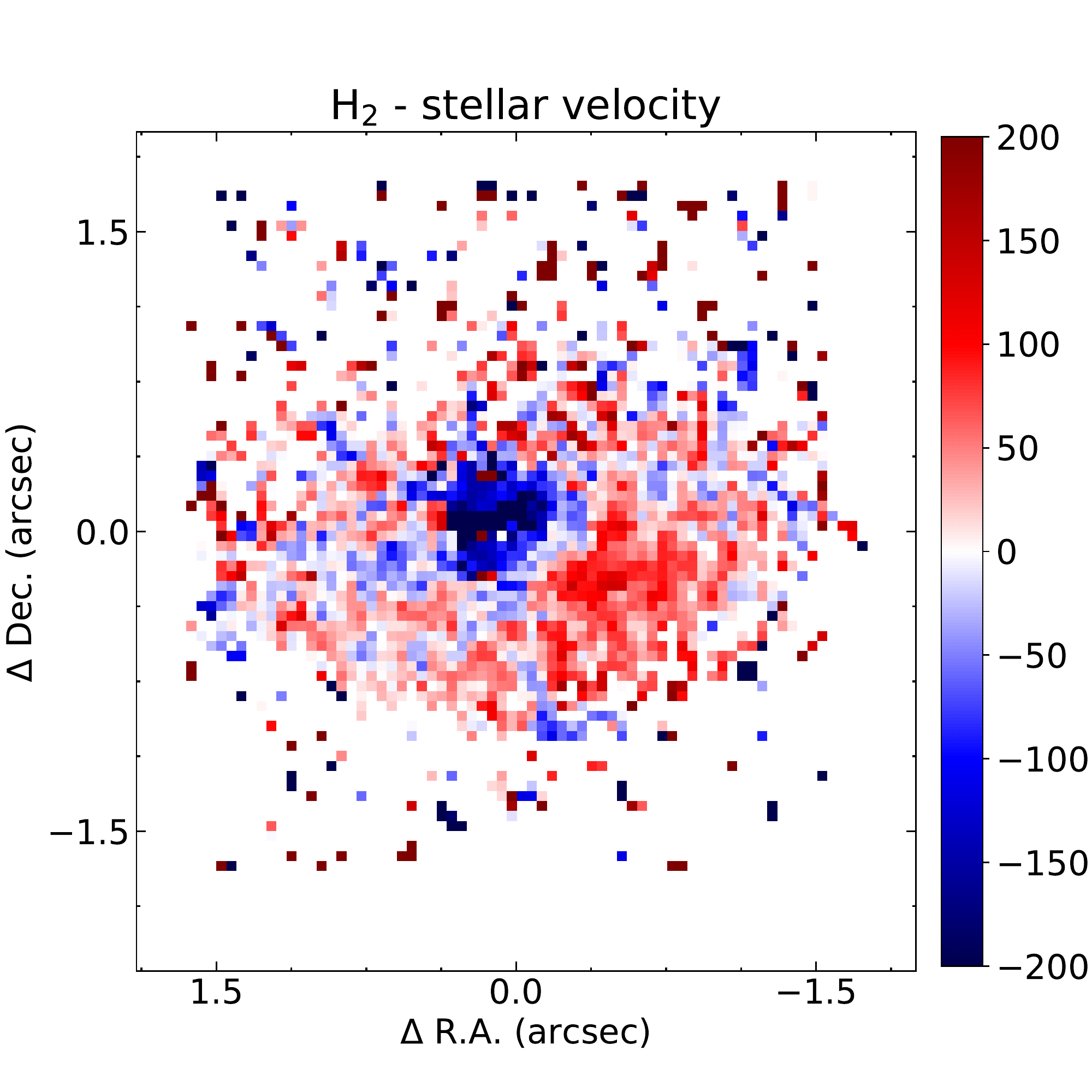}
      \includegraphics[width=7.5cm]{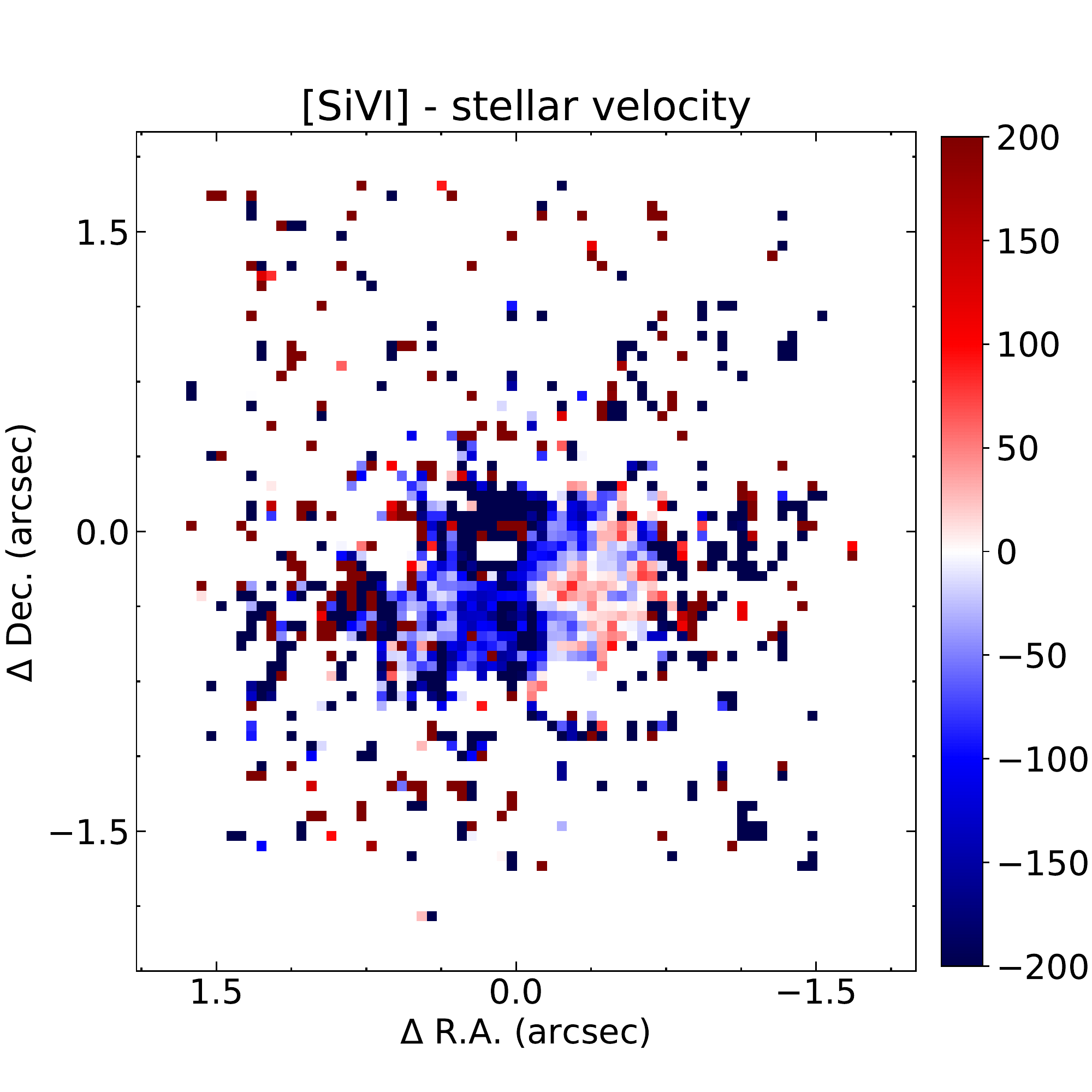}
     \caption{VLT/SINFONI velocity fields of the
     H$_2$ line at $ 2.12\,\mu$m ({\it left panel}) and the [Si\,{\sc
       vi}]$\lambda 1.96\,\mu$m line ({\it right panel}) after subtracting the stellar
       velocity field fitted with {\sc ppxf}. The (0, 0) position
       in all the panels corresponds to 
        the peak of the SINFONI line map. The units of the color bars
        are ${\rm km\,s}^{-1}$.}
              \label{fig:SINFONIminusstars}%
    \end{figure*}

 \section{Morphology and kinematics}\label{sec:morphology}

\subsection{ALMA CO(3-2) and $854\,\mu$m emission}\label{subsec:ALMAmorphology}

Figure~\ref{fig:HSTALMAlarge} (second panel from the top) shows the
ALMA CO(3-2)
   intensity map (moment 0) derived from
the merged configuration data set. It maximizes the best angular
resolution with the best sensitivity of our observations.  The map
displays approximately the full FoV of the
observation along the RA direction. The CO(3-2) 
     intensity map reveals a bright circumnuclear cold molecular gas
     ring with an approximate diameter of  $6-8\arcsec$ or 1.1-1.4\,kpc, 
     containing bright clumps. Beyond the ring, the CO(3-2) emission
     becomes  more diffuse and  covers
     the ALMA FoV along the east-west direction. The presence of this
     ring was already suspected from previous ALMA 
     CO(2-1) images at lower angular resolution
     \citep[see][]{AlonsoHerrero2020}.  For a scale reference, the top
     panel of Fig.~\ref{fig:HSTALMAlarge} displays the
     approximate FoV of the ALMA CO(3-2) image plotted as a red rectangle. The CO(3-2) emission
     clearly traces the prominent dust lane that crosses the (circum)nuclear
     region of the galaxy. In the innermost 4\arcsec, the
     CO(3-2) emission delineates the extinction in the ring derived
     from  the VLT/SINFONI near-IR continuum and the narrow components of hydrogen recombination
     lines \cite[see figure~6 of][]{Smajic2012}.
The CO(3-2) morphology detected in NGC~7172 might correspond to an inner
     Lindblad resonance (ILR) and spiral arms associated to 
     a nuclear bar.
     This CO(3-2) morphology appears to be  similar to that of NGC~613
     \citep{Audibert2019}, although observed at a higher
     inclination. In Sect.~\ref{subsec:ILR} we discuss further
     the ILR.
     
     The ALMA $854\,\mu$m high-angular resolution image from the
     merged configuration data set  (see 
     Fig.~\ref{fig:RINGALMACOcont}, red contours in left panel)
     shows a continuum peak lying at the center
     of the CO(3-2) ring. The measured coordinates are RA =22$^{\rm
       h}$02$^{\rm m}$01.89$^{\rm s}$
     and  Dec =$-$31$^\circ$52'10.49'' (2000), which agree well with those of the ALMA
     1.3\,mm peak emission \citep{AlonsoHerrero2020} and the radio
     emission at 8.4\,GHz \citep{Thean2000}. We thus assumed that this position
     corresponds to the nucleus 
     of the galaxy, that is, it marks the AGN position.
     In the cold molecular gas ring, there is another bright
     $854\,\mu$m source, which is located to the 
     southwest of the nucleus at an approximate projected distance of
    $ 0.5\arcsec$.  There is some faint continuum emission possibly connecting
    the two continuum sources. Other faint continuum peaks are also
    detected. Indeed, the lower angular resolution ALMA $854\,\mu$m
   map from the compact configuration (Fig.~\ref{fig:RINGALMACOcont},
   right panel)  shows both diffuse and compact continuum emission in the cold
    molecular gas ring. From ground-based subarcsecond resolution
    mid-infrared (mid-IR)
     imaging, there is a
     hint of the presence of the two bright sources detected at $854\,\mu$m,
     although the mid-IR emission is 
     dominated by  the AGN \citep[see][]{Roche2007, Asmus2016}. The
     lower angular resolution
     {\it Spitzer}/IRAC $8\,\mu$m image \citep[see][]{GarciaBernete2016} reveals
     emission along the molecular gas ring and again suggests the
     presence of mid-IR emission
     at an orientation of $\sim -120^\circ$, likely arising  
     from the two bright $854\,\mu$m sources
     emission.
     
The map of the CO(3-2) mean velocity (see
Fig.~\ref{fig:HSTALMAlarge}, third 
panel from the top) from  the
merged configuration  shows a  rotation
pattern. There are also clear distorsions in the velocity field, which
might be in part due to the presence of an ILR in NGC~7172. However
the clear departure from rotation in central region along
the minor axis of the galaxy are indicative of the presence of radial
motions. In Sect.~\ref{sec:kinematics}, we perform a detailed modeling
of the CO(3-2) kinematics and quantify the velocity and sign of these
radial motions. The mean velocity dispersion map (see
Fig.~\ref{fig:HSTALMAlarge}, bottom panel) shows a 
complex structure with several regions with velocity dispersions above
$100\,{\rm km\,s}^{-1}$, whereas the most external parts of the disk probed by these
observations show lower values, down to $5-10\,{\rm km\,s}^{-1}$.
     
\begin{figure*}
\vspace{-1cm}
\hspace{1cm}
    \includegraphics[width=16cm]{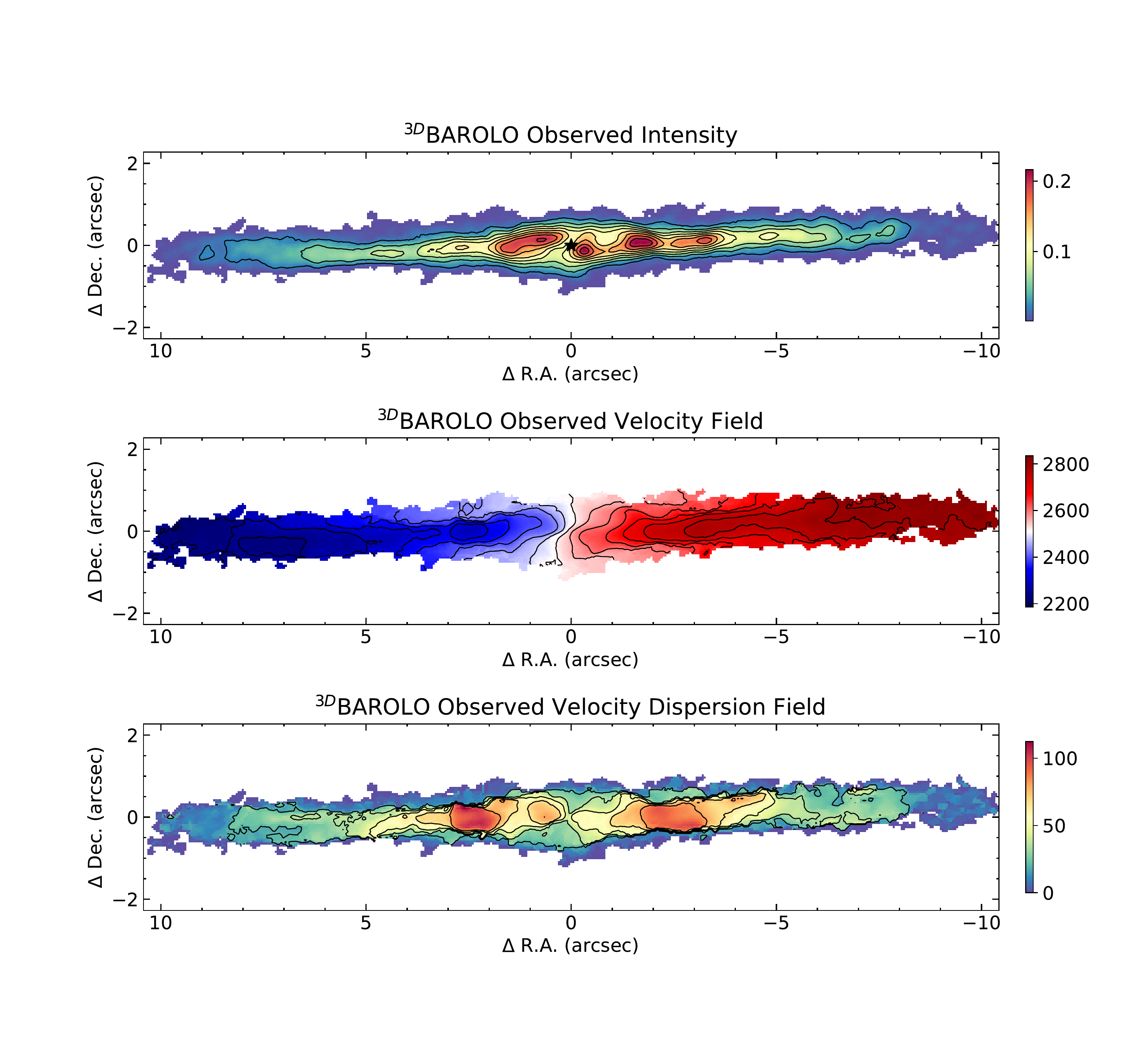}
\vspace{-1.1cm}
\caption{$^{\rm 3D}$BAROLO observed moment maps. From top to bottom, ALMA CO(3-2) maps of the
observed velocity-integrated intensity (moment 0), observed mean
velocity field (moment 1), and
velocity dispersion field (moment 2), produced with the  compact configuration
       observations using $^{\rm 3D}$BAROLO. All the contours
       are on a linear scale. The ALMA beam (not shown) is $0.34\arcsec \times
       0.27\arcsec$ at PA$_{\rm beam}=-74.9^\circ$.
     The units of the velocity and velocity dispersion maps are km
     s$^{-1}$. In the top panel, the star marks
       the approximate location of the AGN.}
              \label{fig:BaroloObservedMoments}%
    \end{figure*}

       \begin{figure*}
\hspace{1cm}
    \includegraphics[width=16cm]{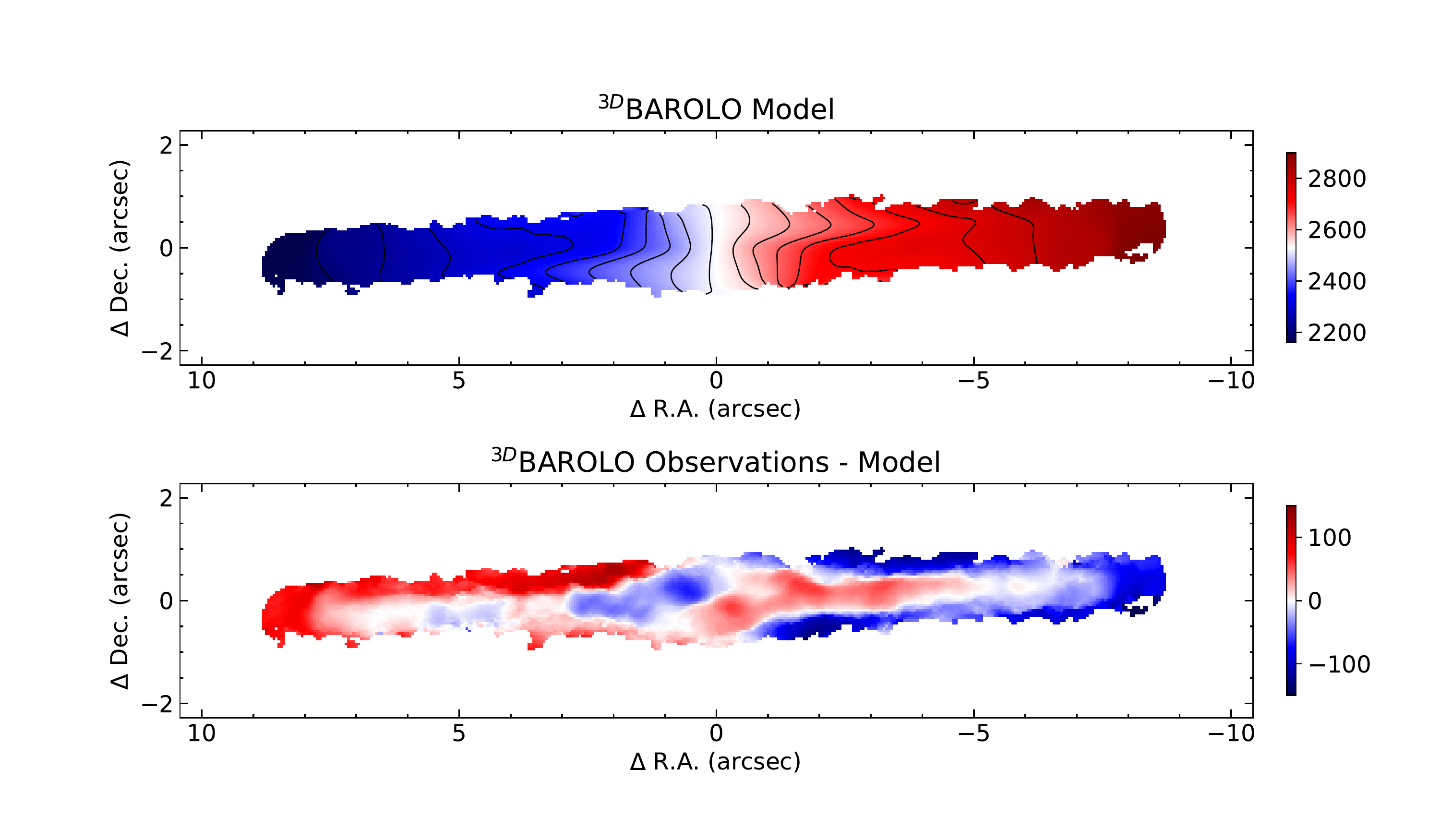}
\vspace{-0.6cm}
     \caption{Velocity map of the $^{\rm 3D}$BAROLO rotating
         disk model without a radial velocity component fitted to the
         ALMA CO(3-2) observations (top) and residual velocity map
         (bottom). The latter was  computed by subtracting the $^{\rm 3D}$BAROLO
       model from the observed moment 1 map. }
              \label{fig:BaroloModelNOVRAD}%
    \end{figure*}

    \subsection{VTL/SINFONI}
     To compare the VLT/SINFONI and ALMA maps, we first aligned them by
     assuming that the peak of the SINFONI emission line and continuum
     maps (the latter not shown here) and the ALMA $854\,\mu$m
     continuum emission coincide. This is justified since
     \cite{Smajic2012} showed that the peaks of the SINFONI near-IR continuum 
     maps mark the position of the nucleus of the galaxy.

     The overall SINFONI H$_2$ $2.12\,\mu$m morphology of the central $\sim
     4\arcsec \times 4\arcsec$ (top-left panel of Fig.~\ref{fig:ALMASINFONI}), which is similar to
       the map from \cite{Smajic2012}, shows two bright sources
       and diffuse emission which extends over larger scales coincident
       with the CO(3-2) ring.  The
       brighest of the two  H$_2$ peaks  arises from the
       nucleus, while the second peak lies close (in projection) and to the
       southwest of  the AGN. The comparison of the SINFONI map with
       the ALMA CO(3-2) (bottom-left panel of Fig.~\ref{fig:ALMASINFONI})
       shows that the secondary H$_2$ peak coincides with one of the
       bright CO(3-2) clumps in the ring of cold molecular
       gas as well as with
       an $854\,\mu$m continuum peak. It is also noticeable that there is bright H$_2$ emission
       at the AGN position, while the CO(3-2) emission is fainter
       there, when
       compared to regions in the cold molecular gas ring. Bright
       H$_2$ and faint CO(3-2) emission at the AGN position has
       been observed in other Seyfert galaxies, for instance,
       NGC~2110 \citep{Rosario2019}. Conversely, we
       do not detect bright H$_2$ emission from other CO(3-2) clumps in the
       ring. Finally, the more diffuse H$_2$ emission appears to be
       filling the region of the inner rim of the CO(3-2) ring but it
       also extends to the ring.

       The extended emission
       from the SINFONI [Si\,{\sc vi}]$\lambda 1.96\,\mu$m line map (top
       right panel of Fig.~\ref{fig:SINFONI}) shows the
       characteristic wide-angle ionization cone
       morphology, similar to that observed in the optical
       emission lines \citep{Thomas2017}. It is also mostly
       extended to the southeast and southwest of the AGN. Likely,
       the other side of the cone is
       obscured by the north rim of the ring, which also marks the
       near side of the galaxy. The projected size of the detected
       emission in this side is approximately $1.5\arcsec \simeq
       270\,$pc. This high-excitation  line only traces the emission in
       the cone closest to the AGN and it is generally between the broad and
       the narrow line region \citep{MuellerSanchez2011}. Indeed,  the optical ionization cone
       traced by the optical [O\,{\sc iii}]$\lambda$5007 emission line is known
       to be extended over much larger scales \citep[see][]{Thomas2017}.
       The second [Si\,{\sc vi}] peak is seen  close in projection to
       the AGN but it appears to be slightly displaced to the east
       with respect to the 
       CO(3-2) and H$_2$ clumps in the ring
       (Fig.~\ref{fig:ALMASINFONI}).

The H$_2$ kinematics (see Fig.~\ref{fig:SINFONI}, middle-left panel)
  shows a rotation pattern with some deviations along the minor axis
  of te galaxy, which
  are similar to those observed in  the CO(3-2) mean velocity
  field. The [Si\,{\sc vi}] velocity
  field (see Fig.~\ref{fig:SINFONI}, middle-right panel) also shows
  an  apparent rotation pattern, although it can 
  be seen that it is not exactly at the same orientation and velocity
  values as that of the H$_2$ mean velocity field (see below).
  The presence of rotational + outflow motions in the coronal
  [Si\,{\sc vi}]  line  has been inferred in other AGNs
  \citep{MuellerSanchez2011, RamosAlmeida2017}.  The velocity dispersion maps of
  the two lines are
  quite different. The  H$_2$ line (Fig.~\ref{fig:SINFONI}, bottom-left
   panel) shows, except at the AGN position,  a mostly uniform
  distribution of with $\sigma({\rm H}_2)\simeq100\,{\rm km\,s}^{-1}$. These
  values are  slightly higher than those measured in the central
  regions of other Seyfert
  galaxies \citep{Hicks2009}. The [Si\,{\sc vi}] velocity dispersion map
  (Fig.~\ref{fig:SINFONI}, bottom-right panel), on the other hand,
  displays an interesting structure. In general, the measured values
  of the velocity dispersion indicate the presence of relatively broad
  lines near the AGN position with $\sigma({\rm [SiVI]})=400-600\,{\rm km\, s}^{-1}$
  (FWHM$=900-1400\,{\rm km\,s}^{-1}$),
  which are indicative of the presence of an ionized 
 gas outflow (see below). The
  exception is the region where the AGN wind appears to have impacted
  the disk of the galaxy, where the velocity dispersions are markedly
  decreased.

 To evaluate the presence of noncircular motions in the H$_2$ and
 [Si\,{\sc vi}] velocity fields, we subtracted the stellar velocity
 field fitted with {\sc ppxf}. The H$_2$ velocity residual map shows
 (Fig.~\ref{fig:SINFONIminusstars}, left panel)  red
 and blue excesses to the northeast and southwest, respectively. If
 these motions are taking place in the plane of the galaxy, they would
 indicate the presence of a molecular outflow in the galaxy
 disk, as the near side of the galaxy is to the north. Given the higher velocity resolution of the ALMA
 CO(3-2) observations, in
 Sect.~\ref{sec:kinematics} we  model  the kinematics of this
 molecular gas
 transition to separate the rotation from the noncircular motions. The [Si\,{\sc
   vi}] velocity residual map (see Fig.~\ref{fig:SINFONIminusstars},
 right panel), on the other hand, displays mostly 
 blueshifted velocities of a few hundred ${\rm km\,s}^{-1}$ in the southern
 part of the ionization cone out to  $\sim 200\,$pc (in projection)
 from the AGN. These noncircular motions together with
 the large FWHM of the lines fitted in the same region
 can be interpreted as an ionized gas outflow outside the plane
 of the galaxy. It likely traces the inner part of the larger scale outflow
 detected in the
 [O\,{\sc iii}]$\lambda$5007 emission line
 \citep{Thomas2017, Davies2020}. The milder red residual velocities of
 the [Si\,{\sc vi}] line to the
 southwest of the AGN 
 are associated with the brightening of the line described above,
 which probably indicates that the AGN is illuminating a clump of gas
 in the disk of the galaxy rather than in the ionized gas outflow
 region. The redshifted velocities might indicate outflowing motions
 in the disk of the galaxy.

\section{Modeling of the ALMA CO(3-2)
  kinematics}\label{sec:kinematics}

In this section we model the ALMA compact configuration
CO(3-2) kinematics of NGC~7172 with the 
$^{\rm 3D}$BAROLO code \citep{DiTeodoro2015}. The angular resolution of
these observations (see Table~\ref{tab:ALMAobservations}) allows to
resolve the kinematics on physical scales of $\sim 54\,$pc.  
This code fits simple rotating disk models
using 3D tilted rings and can be used with a variety of emission line
data, including ALMA  data cubes.

\begin{figure}
     \includegraphics[width=9cm]{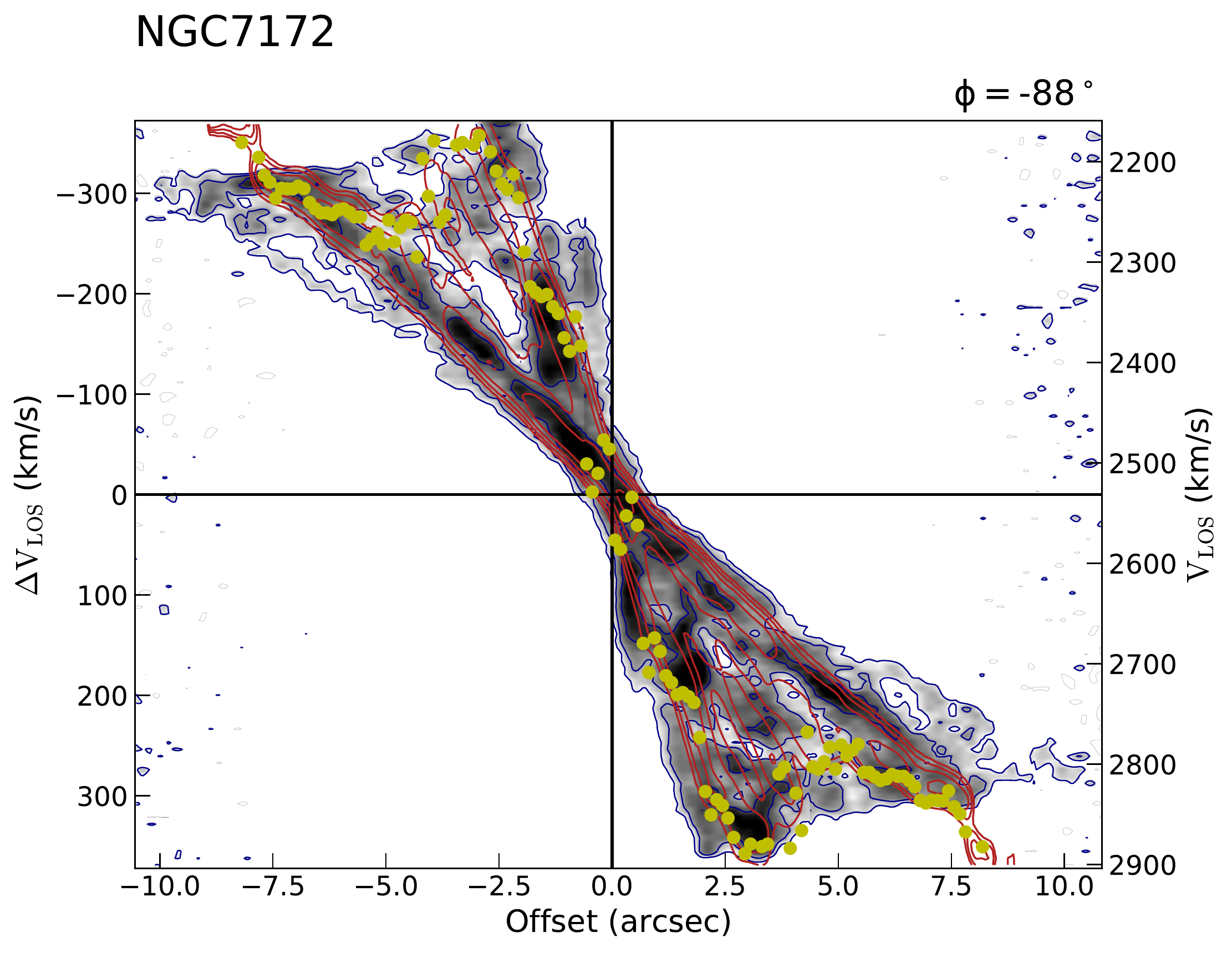}
     \includegraphics[width=9cm]{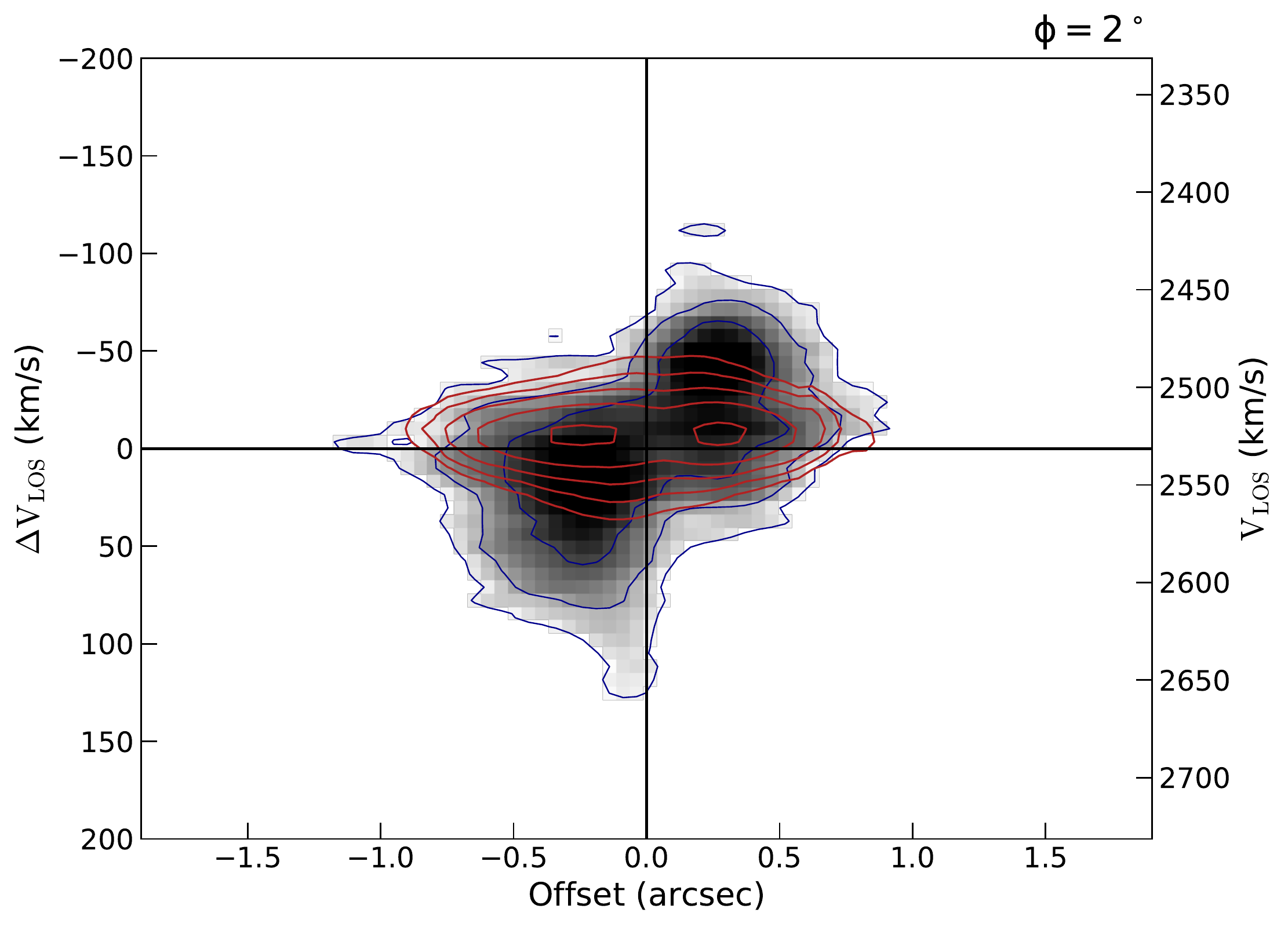}
     \caption{ALMA CO(3-2) p-v diagrams generated with $^{\rm 3D}$BAROLO along
       the kinematic major ({\it top panel}) and minor ({\it bottom panel}) 
       axes. The gray scale and blue contours are the ALMA CO(3-2) observations,
       while the red contours are the $^{\rm 3D}$BAROLO rotating disk
       model. The yellow dots are the fitted rotation curve.}
              \label{fig:BaroloPVNOVRAD}%
\end{figure}

\subsection{Rotating disk}\label{subsec:Barolodisk}
 $^{\rm 3D}$BAROLO can fit a
number of free parameters for a rotating disk model, including the
kinematic center, the systemic velocity, the disk inclination ($i$) and
position angle (PA) of the kinematic major axis, the scale height of the disk, the circular
velocity, and the gas velocity dispersion ($\sigma_{\rm gas}$). It
  also takes into account effects related to the beam smearing in the
  definition of the best-fit model
  for a given spatial resolution. $^{\rm
  3D}$BAROLO also produces 
maps of the 
observed velocity-integrated intensity (moment 0), a
map of the observed mean velocity field (moment 1) and
velocity dispersion field (moment 2), as well as position-velocity (p-v)
diagrams along the kinematic major and minor 
axes. Before we started the fits, we trimmed the original compact
configuration data cube to the FoV with detected emission,  $\sim 20.9\arcsec \times
4.6\arcsec$.

Figure~\ref{fig:BaroloObservedMoments} shows the ALMA 
CO(3-2) observed moment maps  of NGC~7172. The intensity, velocity,
and velocity dispersion
maps are
similar to those in Fig.~\ref{fig:HSTALMAlarge}, considering the
difference in angular resolution. The CO(3-2) mean velocity field
shows a general rotation pattern. However, some important distorsions
are readily seen in the molecular gas ring, which are accompanied by
the presence of high velocity dispersion values, especially along the
east-west direction in the disk of the galaxy.  The CO(3-2) velocity
dispersions show relatively large values of up to $\sim 100\,{\rm
  km\,s}^{-1}$ in the ring, whereas in the
outer parts of the disk mapped with ALMA the velocity dispersions
decrease to $\sim 5\,{\rm km\,s}^{-1}$.

\begin{figure*}
\hspace{1cm}
    \includegraphics[width=16cm]{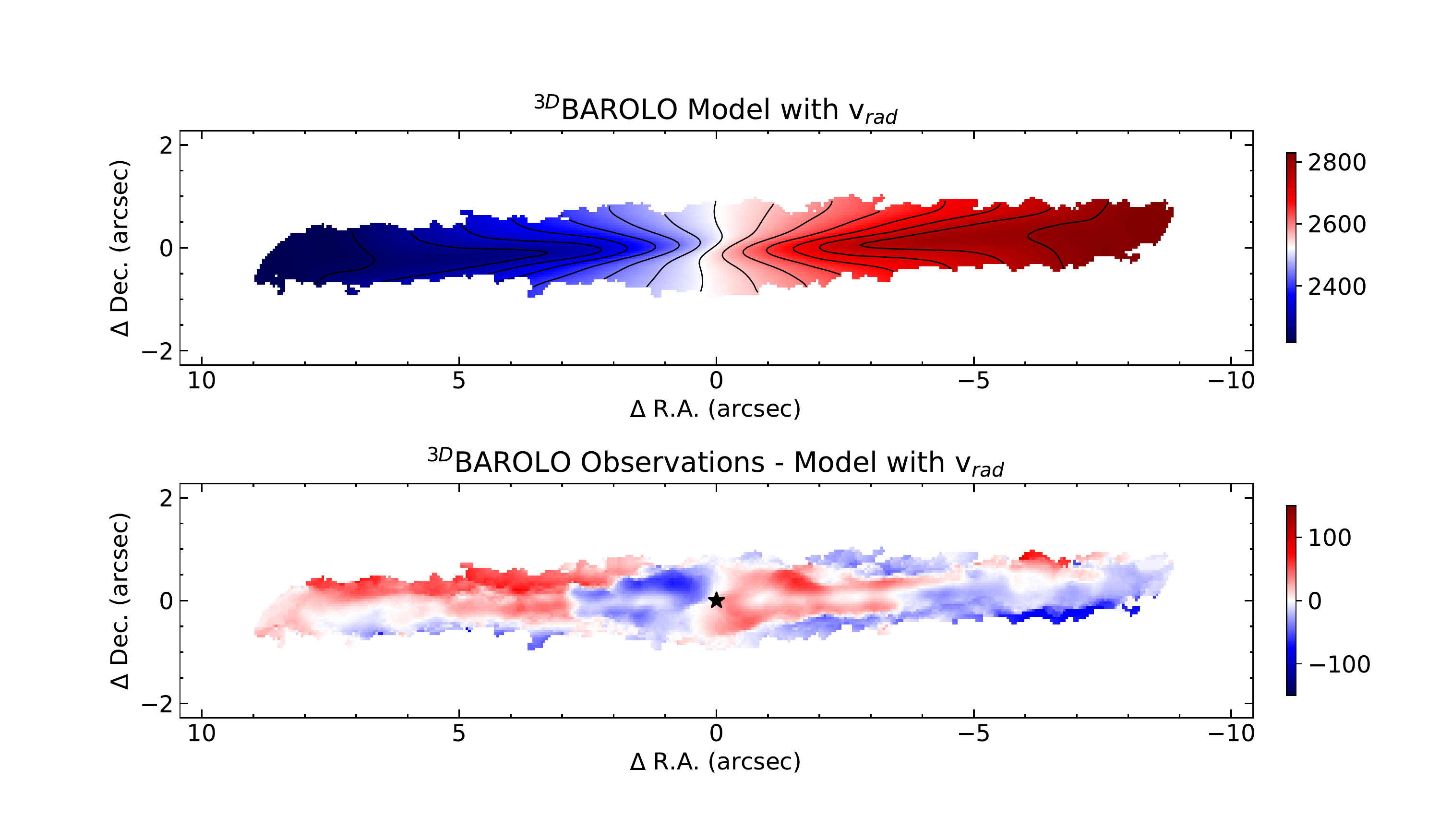}
\vspace{-0.6cm}
     \caption{ Velocity map of the $^{\rm 3D}$BAROLO rotating disk model with a radial velocity
         component and residual velocity map. Panels are the same as in
         Fig.~\ref{fig:BaroloModelNOVRAD}.  The
       star symbol shows 
       the approximate location of the AGN.}
              \label{fig:BaroloModelVRAD}%
    \end{figure*}

      \begin{figure}
     \includegraphics[width=9cm]{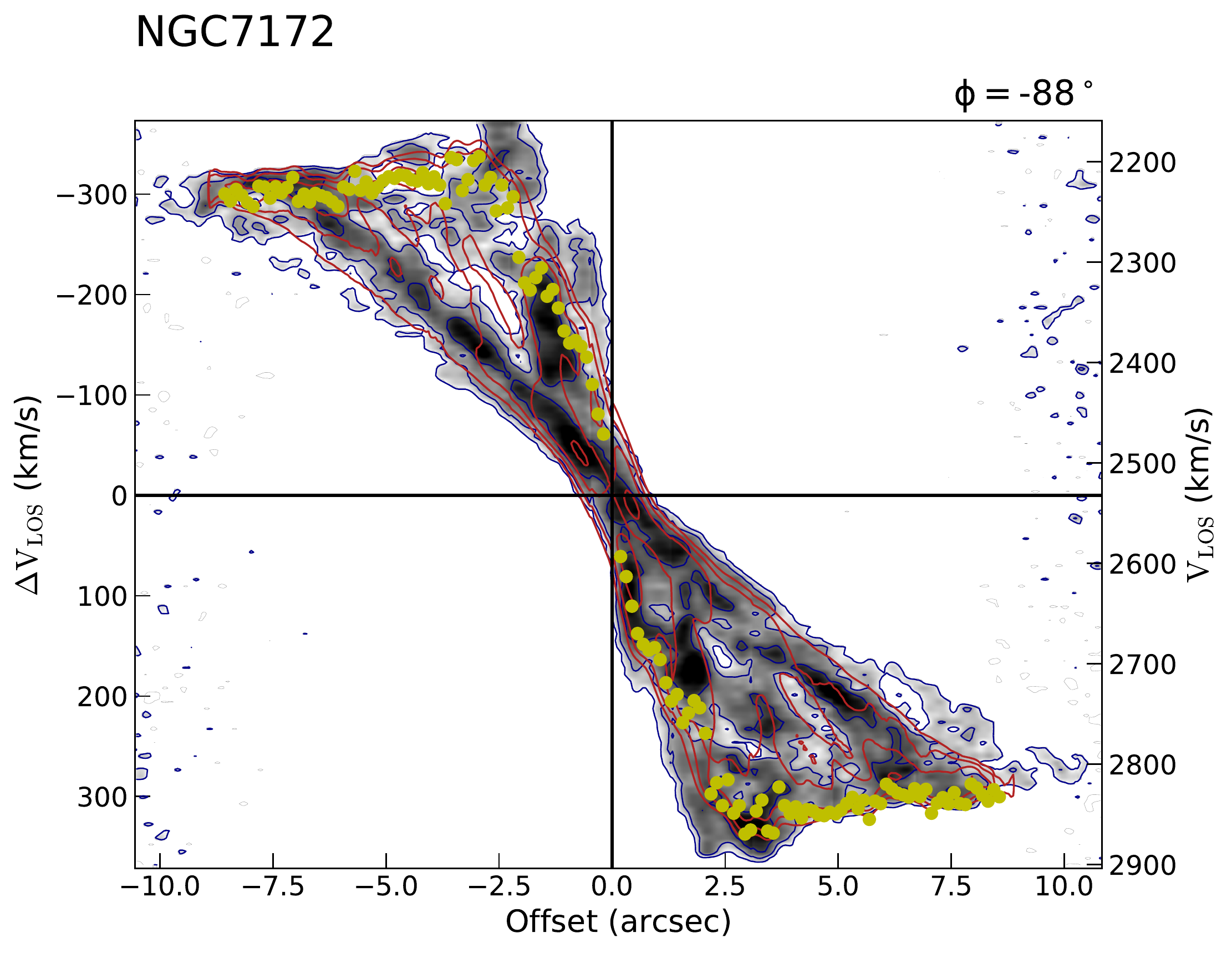}
     \includegraphics[width=9cm]{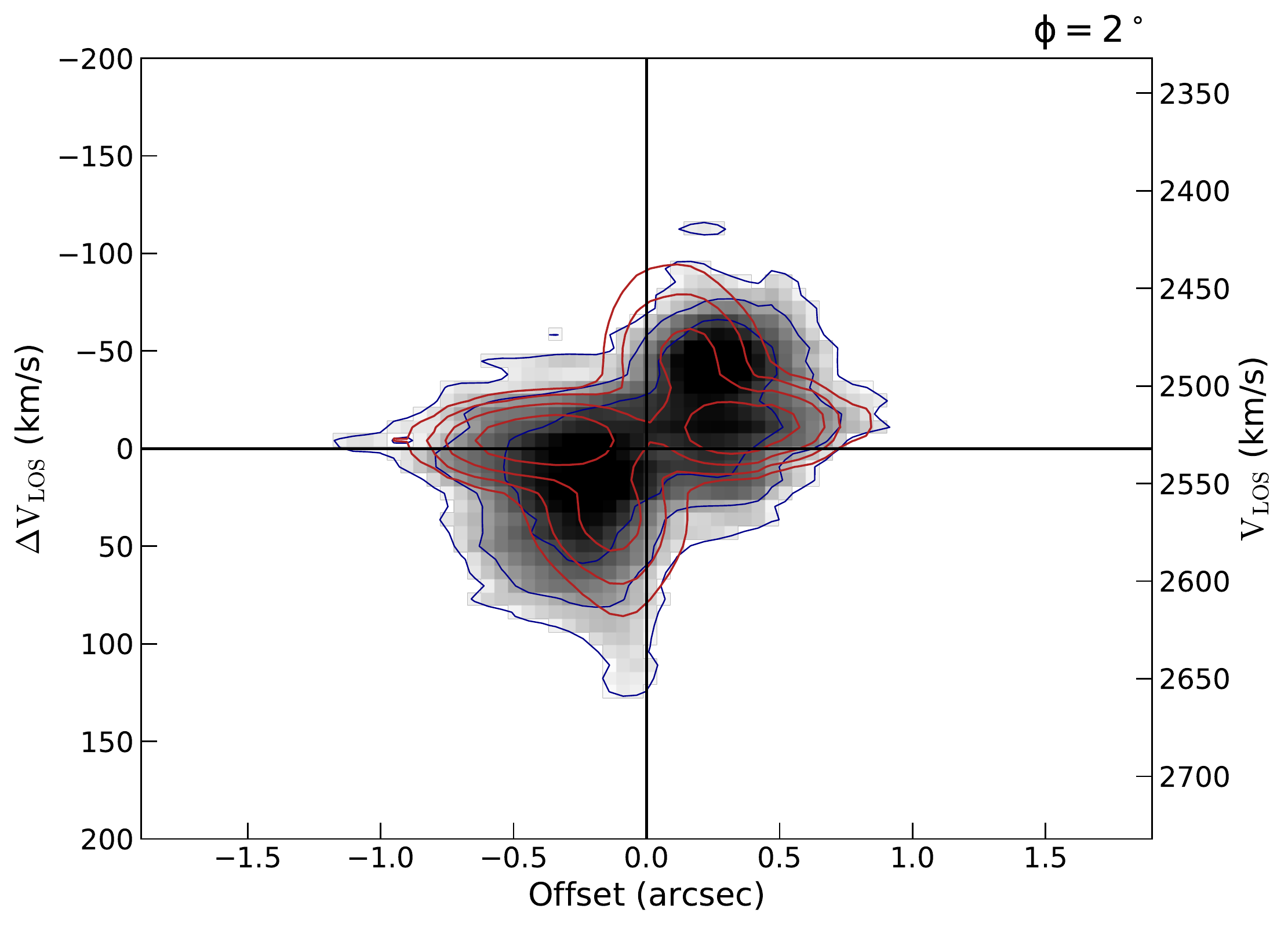}
     \caption{ALMA CO(3-2) p-v diagrams generated with $^{\rm
           3D}$BAROLO. Panels, colors, and lines are the same as Fig.~\ref{fig:BaroloPVNOVRAD}, but for the  $^{\rm
         3D}$BAROLO disk model with a radial velocity component. The
       yellow dots in the top panel are the fixed rotation curve used for this model.}
              \label{fig:BaroloPVVRAD}%
    \end{figure}

For all the   $^{\rm 3D}$BAROLO fits, we fixed the kinematic center 
at the position of the $854\,\mu$m continuum peak (see Sect.~\ref{subsec:ALMAmorphology}) and 
used a disk Gaussian scale height of $0.05\arcsec$ ($\simeq 10$\,pc, equivalent
to a 24\,pc FWHM at the assumed
distance). Since NGC~7172 is seen at a high inclination (see below), this
  relatively thin disk is needed to avoid mixing the different emitting regions in
  our line of sight. We took a radial separation between the rings of 
$0.125\arcsec$ and a total of 70 rings, thus reaching out to radial
distances of approximately $8.75\arcsec$ ($\simeq 1.6\,$kpc).
Given the high inclination of
NGC~7172, we first measured it from the apparent relative size of the
cold molecular gas ring, assuming that it has a circular shape.
We obtained $i=85^\circ$, which we allowed to vary
by a few degrees during the fit. The observed PA of the CO(3-2) kinematics appears to be close to the
east-west direction, as found for the optical emission lines
\citep{Thomas2017}. We thus allowed our fit to vary this parameter by $10^\circ$. 
 During the first step, we fit five parameters, namely, $i$, PA,
 systemic velocity, circular velocity, and $\sigma_{\rm gas}$. For the
 last parameter we imposed a minimum value of $5\,{\rm km\,s}^{-1}$,
 as seen in the observations.

 We derived a systemic velocity of 2532\,km
 s$^{-1}$ and an average value of the inclination of $ i = 88^\circ$. 
 Figure~\ref{fig:BaroloModelNOVRAD} shows the  best-fit
 model CO(3-2) mean velocity field (top
   panel) and the CO(3-2) mean velocity residuals (bottom panel), which
   were computed by subtracting
   the model velocity from the observed velocity field. There are
   velocity residuals of a few tens to  $\simeq 100\,{\rm
     km\,s}^{-1}$, in the central region and in the outer parts of the
   galaxy disk.
To understand better the mean velocity residuals, we produced p-v
diagrams along the kinematic major and minor axes of the galaxy.
 The observed p-v diagram along the kinematic major axis
   (Fig.~\ref{fig:BaroloPVNOVRAD}, top panel, gray scale map and blue contours) has a tilted X-shape,
   which is the result of beam smearing and the ring-like structure of the
   highly inclined disk of NGC~7172.
The $^{\rm 3D}$BAROLO rotating disk model (red contours in
Fig.~\ref{fig:BaroloPVNOVRAD}) produces a reasonably good fit
to this p-v diagram.  
   However, the fitted rotation
   curve (yellow dots in the figure) does not reach the terminal velocity at radial distances
   $r\sim 3-6\arcsec$ and is not well fitted in the innermost
   regions.
   Both are in part responsible for the large
   velocity residuals seen in Fig.~\ref{fig:BaroloModelNOVRAD}
   (bottom panel). Along the kinematic minor axis 
(Fig.~\ref{fig:BaroloPVNOVRAD},  bottom panel), there are clear
noncircular motions in the central $2\arcsec$ that we address in the see next section.

\subsection{Rotating disk with radial velocity}\label{subsec:BarolodiskVRAD}
    
Within the
approximate inner $2\arcsec$, the minor axis p-v diagram
shows redshifted motions to the
south of the AGN and blueshifted to the north
(see Fig.~\ref{fig:BaroloPVNOVRAD}, bottom panel). Since the north
is the near side of the galaxy (see the top panel of Fig.~\ref{fig:HSTALMAlarge}), this suggests the 
presence of an outflow in the cold molecular gas ring, if these motions are taking place in the plane
of the galaxy. We thus run another   $^{\rm 3D}$BAROLO model including a
radial velocity  ($v_{\rm rad}$) component. We fixed the inclination to the average
value from the previous run and the derived systemic velocity. We also
fixed the rotation curve to agree better
with the observed p-v diagram along the major axis. We
therefore left PA, $\sigma_{\rm gas}$, and $v_{\rm
  rad}$ as free parameters.  

\begin{figure}
\vspace{-0.1cm}
  \hspace{-0.1cm}
     \includegraphics[width=9cm]{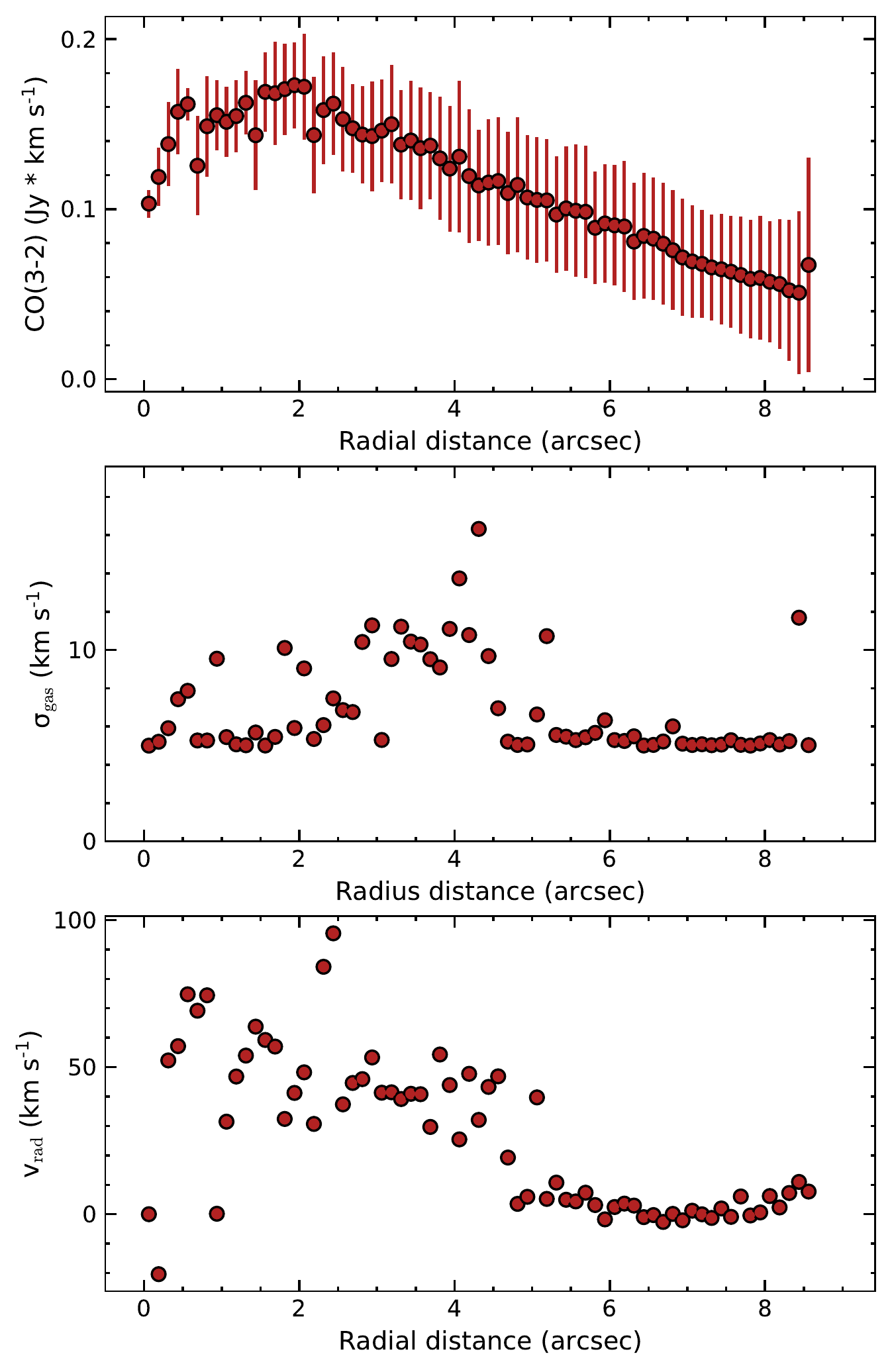}
\vspace{-0.5cm}
     \caption{Radial profiles  of the ALMA CO(3-2) median 
  intensity ({\it top panel}), gas velocity
       dispersion ({\it middle  panel}), and radial velocity ({\it
         bottom panel}) fitted with the $^{\rm 3D}$BAROLO model in
       Sect.~\ref{subsec:BarolodiskVRAD}.
       As explained
       by \cite{GarciaBurillo2019}, the vertical error bars in the
       integrated CO(3-2) emission reflect the deviations from symmetry in the gas distribution.}
              \label{fig:radialdistributions}%
    \end{figure}

We show the
resulting mean velocity model and velocity residual map in
Fig.~\ref{fig:BaroloModelVRAD} (top and bottom panels, respectively).
It is immediately clear from the comparison with
Fig.~\ref{fig:BaroloModelNOVRAD}, that the $^{\rm 3D}$BAROLO model
with the fixed rotation curve and
the radial velocity component produces a better fit to the
observations. The effect of including the latter is appreciated in  
the CO(3-2) p-v diagram along the kinematic minor
axis (see Fig.~\ref{fig:BaroloPVVRAD}).
However, even when including radial
velocities, the model does not reach the observed velocity of nearly $+150\,{\rm km\,s}^{-1}$ to
the south, close to the AGN position. The maximum fitted
radial velocities are  $80-100\,{\rm km\,s}^{-1}$ (see the
bottom panel of Fig.~\ref{fig:radialdistributions}), progressively
decreasing to $\simeq 20-30\,{\rm
  km\,s}^{-1}$ out to a  radial distance of $\simeq 4.5\arcsec$. This deceleration might be
due to the AGN wind encountering large amounts of molecular gas.
The bright peaks of H$_2$, CO(3-2), and $854\,\mu$m
emission in the ring (see Sect.~\ref{sec:morphology})
 might be caused by this impact.
Within the gas ring, the gas velocity
dispersion takes values of approximately $\sigma_{\rm gas} \simeq 10\,{\rm
  km\,s}^{-1}$, except in the region with the largest
radial velocity values  (see the 
middle panel of Fig.~\ref{fig:radialdistributions}) where the gas
dispersion is slightly lower. At
$r > 4.5\arcsec$, the velocity dispersion goes down to the minimum
value imposed in our fit ($\sigma_{\rm gas}=5\,{\rm km\,s}^{-1}$).

Despite having included a radial velocity component in the $^{\rm
  3D}$BAROLO model, there are still some CO(3-2)
mean velocity residuals in some regions of the map
(Fig.~\ref{fig:BaroloModelVRAD}, bottom panel). Some are beyond the expected
virial motions around rotation of typically a few tens of ${\rm km\,s}^{-1}$. In the central
$2\arcsec$ of NGC~7172, there are two regions with moderate velocity residuals  of the
order of $50\,{\rm km\,s}^{-1}$, which are redshifted to the southwest of the AGN and blueshifted
to the northeast. They appear to be coincident with those regions where the innermost part
of AGN wind, traced by the VLT/SINFONI [Si\,{\sc vi}] emission (see
left panels of Fig.~\ref{fig:ALMASINFONI}), impacts the inner part of the cold molecular
gas ring. Even though at the radial distances of these regions we
fit the largest values of
the radial velocities (bottom panel of
Fig.~\ref{fig:radialdistributions}), the residuals mean that this
azimuthally averaged model cannot explain completely the observed kinematics. In fact, these distinct velocity
residuals also define kinematically the  expanding molecular ring with a
diameter of $\sim 6\arcsec$ or approximately $1.1$\,kpc.
The residual mean velocities are not only confined to these regions,
but there are some residuals beyond the molecular ring, mostly to the
northeast.

We note that the $^{\rm 3D}$BAROLO tool does not take into account the
potential contribution from other 
terms of the Fourier decomposition
of noncircular motions. These include tangental motions or higher
order terms that could  improve the quality of the fit
\citep{Schoenmakers1997}. The missing terms, which can capture better
the density wave streaming motions produced by a stellar bar, are in
principle more suitably 
accounted for by tools like {\tt kinemetry}
\citep{Krajnovic2006}. This is particularly clear in  those cases
where the mean velocity fields provide a realistic description of the gas
motions, that is, in galaxy disks seen at favorably low
inclinations\footnote{We nevertheless note that {\tt kinemetry} would
  be of 
  limited use to fit the 
observed mean velocity field of NGC~7172 due to its high inclination.}.
\cite{GarciaBurillo2019} benchmarked the solutions found by  $^{\rm 3D}$BAROLO
against the output provided by {\tt kinemetry} for the disk of  NGC~1068, which is seen at $i\simeq41^\circ$, and  concluded that there is an
 overall excellent agreement in the sign and magnitude of the $v_{\rm
   rad}$ radial profiles derived by  the two
 modeling tools,  in spite of their different
 working approaches.  

In this context, it is worth recalling that in all realistic
gas configurations for the gas response to a stellar bar  inside its
corotation (see Sect.~\ref{subsec:ILR}, for a discussion of an
alternative scenario),  
the average value of the $v_{\rm rad}$ term should reflect 
prevalent inward radial motions  along the bar leading edges 
and down to the ILR ring. These motions are due to the orbit crowding
of precessing individual elliptical orbits \citep[see   
e.g.,][]{Wong2004, GarciaBurillo2014, GarciaBurillo2019}.
In particular, for the case of a spiral-like pseudo-ring formed outside and down to the ILR, gas motions
should be dominated either by inflow or, in the case of a circular
ring  by no-flow at all \citep[see  
e.g.,][]{Wong2004}. Other orbit configurations not contemplated in the models published by \citet{Wong2004} cannot
be formally excluded, like those related to the presence of a nuclear
bar or episodic leading gas spirals. However, there is no
evidence for any of these in NGC~7172.  Moreover, the  fact that the sign
of $<$$v_{\rm rad}$$>$$(r)$ in the central $\simeq700$~pc of NGC~7172 is
reversed relative to the basic theoretical expectation
(i.e., no flow $<$$v_{\rm rad}>$$=0$ or inflow based on the sign of $<$$v_{\rm rad}$$>$)
suggests that there is a (true) radial (out)flow component superposed  to the usual terms
that could be potentially attributed to bar-driven noncircular motions in these regions.
This is similar to the scenario described by \citet{GarciaBurillo2014,
   GarciaBurillo2019} for NGC~1068,
where the presence of the molecular outflow was unambiguously  detected by
a clear reversal of the $s1$ term in the Fourier decomposition of the velocity field of the gas in the
inner region of the galaxy disk. Both $^{\rm 3D}$BAROLO  and
the Fourier decomposition done by {\tt kinemetry} capture this signature of the
outflow, that is, the sign reversal of radial motions \citep{GarciaBurillo2014,
   GarciaBurillo2019}.

\subsection{Signatures of an ILR in the observed CO(3-2) kinematics}\label{subsec:ILR}
In Sect.~\ref{subsec:ALMAmorphology} we proposed that the conspicuous
ring morphology displayed by the CO(3--2) merged configuration emission 
might be explained in terms of the canonical gas response to an ILR
 of a stellar bar \citep{Buta1996}. The existence of a
large-scale  stellar bar in NGC~7172 is still debated. Based on the
modeling of the near-IR 2MASS
isophotes, \cite{MenendezDelmestre2007} reported the existence of a large-scale bar
(PA=$96^\circ$ and a semi-major axis of $a_{\rm
  bar}=45\arcsec \simeq 8\,$kpc), while 
\cite{Lin2018}  did not require one.
However, the noncircular motions of the molecular gas revealed by
the CO data in the central $\simeq1-1.5$~kpc
of the galaxy disk might be reflecting the influence on the gas kinematics  of a (not yet
observed) nuclear
bar located  within the large-scale stellar bar.   
This scenario is explored in detail in the numerical simulations of 
Appendix~\ref{sec:appendix}. To explain the observed CO(3-2) molecular ring size, the
radius of this {\it putative} nuclear bar needs 
to be significantly smaller, of the order of $r=2$\,kpc,  than that of
\cite{MenendezDelmestre2007}.

As illustrated in Appendix~\ref{sec:appendix}, the observed S-shape of the CO(3-2) mean
velocity field and the outflow signature along the minor axis  
could be reproduced reasonably well by selecting a favorable phase angle of a
nuclear bar in the plane of the galaxy (see Figs.~\ref{fig:mom0} and \ref{fig:pvmai}). 
These noncircular motions reflect the projection of elliptical
streamlines corresponding to the $x_{\rm 1}$ orbits  
building  the potential of the nuclear bar. In particular, a good
correspondence with the data requires that  the  $2$\,kpc-semi-major
axis nuclear bar  is seen  at a moderate inclination ($i=60^\circ$).
This is nevertheless in stark contrast with the high value of 
the inclination needed to fit the CO(3-2) kinematics with $^{\rm
  3D}$BAROLO. Furthermore, in these simulations the observed CO(3-2) morphology 
appears as a result of the pile-up of gas in a highly elliptical ring
at the location of a single ILR. The nuclear bar in these simulations 
is a ``fast'' $m=2$ mode. As a result,  there is little room for
$x_{\rm 2}$ orbits,  which are only present when there are two distinct ILRs. 
This physically possible, yet limit case, configuration has two
implications.  First, the gas piles up at
a highly elliptical ILR ring and, secondly, the $x_{\rm 1}$
orbits do not precess into $x_{\rm 2}$ orbits and, as a consequence,  the formation of the canonical 
bar leading edges feature ---favoring the scenario described in
Sect.~\ref{subsec:BarolodiskVRAD}---  would be thwarted.

While the configuration described above and in
Appendix~\ref{sec:appendix} for a putative nuclear bar is
an alternative interpretation for the observations, the lack
of any direct evidence of its existence  
disfavors this scenario versus the AGN-driven outflow scenario
we discuss in Sect.~\ref{sec:discussion}.

\begin{table}
\caption{Nuclear and circumnuclear cold molecular gas properties.}\label{tab:coldmoleculargas}
\centering
\begin{tabular}{lcccc}
\hline\hline
Region& $r$& $i$ & CO(3-2) flux & $M_{\rm gas}$\\
&  (pc) & ($^\circ$) & (Jy kms$^{-1}$) & ($M_\odot$)\\
  \hline
  Torus & 32 & 67 & 1.4 & $8.2\times 10^5$\\
  Nuclear & 50 & 85 & 0.9 & $7.2\times 10^5$\\
  Circumnuclear & 200 & 85 & 21 & $1.7\times 10^7$\\
  Ring & 720 & 85 & 270 &  $6.2\times 10^8$\\
  \hline
\end{tabular}
\tablefoot{We used different brightness temperature ratios
  for the torus and the other regions, namely, $T_{\rm
    B}$CO(3-2)/$T_{\rm B}$CO(1-0)$=2.9$ for the torus, $T_{\rm
    B}$CO(3-2)/$T_{\rm B}$CO(1-0)$=2.0$ for the nuclear and
  circumnuclear regions, and $T_{\rm B}$CO(3-2)/$T_{\rm
    B}$CO(1-0)$=0.7$ for the ring. See text for details. }
\end{table}

    \section{The torus of NGC~7172}\label{sec:torus}
    
The first ALMA detection of an AGN torus was for the Seyfert galaxy
NGC~1068 \citep[e.g.,][]{GarciaBurillo2016, Gallimore2016, Imanishi2018,
  GarciaBurillo2019, Impellizzeri2019}. The NGC~1068 torus is a multiphase structure
with diameters  in the 
$7-40\,$pc range, depending on the molecular transition and the
cold dust continuum. In other nearby AGNs, ALMA high-angular resolution
(typically less than $0.1\arcsec$) observations of the cold dust and
molecular gas have proven to  be the most efficient way to resolve
molecular dusty
tori. In the approximately twenty detections reported so far,
the tori/disks are found to be  relatively  large, with diameters $20-130\,$pc, and
masses in the $10^5-10^7\,M_\odot$ range in Seyfert galaxies \citep{Izumi2018, AlonsoHerrero2018,
AlonsoHerrero2019, GarciaBurillo2021}  and low-luminosity
AGNs  \citep{Combes2019}. Moreover, the molecular gas 
column densities at the AGN location appear to be broadly correlated with the X-ray absorptions
$N_H$, but  physical resolutions of the order of  10\,pc are
needed to reach the relevant scales responsible for 
AGN obscuration \citep{AlonsoHerrero2018, GarciaBurillo2021}.

\begin{figure}
\vspace{-1cm}
  \hspace{-0.75cm}
     \includegraphics[width=18cm]{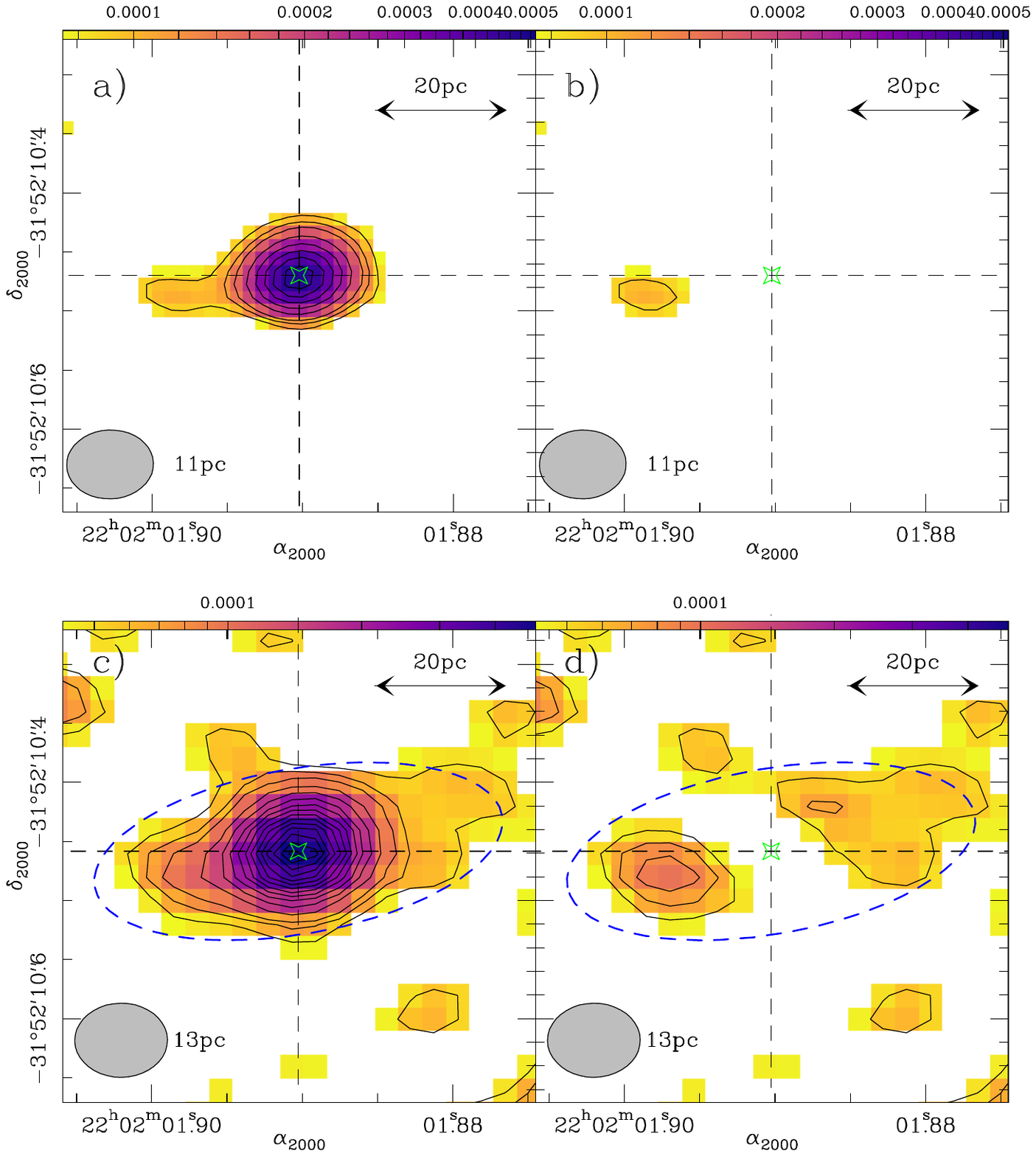}
\vspace{-1.5cm}
\caption{Central $0.4\arcsec \times 0.4\arcsec$ region showing
  the ALMA $854\,\mu$m emission. {\it Top panels:} Observed ({\it left panel}) and point-source
  subtracted ({\it right panel}) images from the extended
  configuration data set. {\it Lower panels:} Same as the {\it upper panels}, but
  for the merged configuration. The dashed blue ellipse marks the
approximate extent of the torus.  In all the panels, the
first contour is at $3 \times$ rms. The gray ellipses represent the corresponding
observation beams.}
              \label{fig:torus}%
    \end{figure}

The left panels of Fig. ~\ref{fig:torus} show the observed $854\,\mu$m
continuum of the central $0.4\arcsec \times 0.4\arcsec$ (approximately
$72\,{\rm pc} \times 72\,{\rm pc}$) for the observations with the
highest angular resolution (extended configuration, top panel) and the
intermediate resolution (merged configuration, bottom panel). The
corresponding physical resolutions are approximately 11\,pc and
13\,pc, respectively. From the
comparison with the beam of the 
merged configuration observations, it is immediately clear that the continuum
emission is extended. In the bottom panels of this figure we marked
the extent of the torus  with a blue dashed ellipse, which traces
approximately the 
$3\times$rms contour (see below). The orientation of this extended emission
appears to be nearly perpendicular to the [Si\,{\sc vi}]$\lambda 1.96\,\mu$m emission,
which indicates that it is likely tracing equatorial emission in the
dusty torus.
 The aspect ratio of the $854\,\mu$m emission
(before the point-source subtraction, see below) implies a torus inclination of
approximately $i_{\rm torus}=67^\circ$. In fact, this needs to be
taken as an upper limit because the torus is likely to have some
thickness. Nevertheless, this estimate is in good agreement within
intermediate inclinations inferred
 from the fitting of the infrared and X-ray emission with torus models
 for NGC~7172
\citep{RamosAlmeida2009, RamosAlmeida2011, AlonsoHerrero2011, Ichikawa2015, Vasylenko2018, Tanimoto2020}. The derived torus radius is
approximately $0.18\arcsec \sim 32$\,pc,
which is within the range observed for other Seyfert galaxies and
low-luminosity AGNs \citep{AlonsoHerrero2018, Combes2019,
  GarciaBurillo2021}.

Nevertheless, as found in other Seyfert galaxies, it is
likely that  even at $854\,\mu$m there is some contribution
from synchrotron emission \citep{Pasetto2019, AlonsoHerrero2019,
  GarciaBurillo2019, 
  GarciaBurillo2021}. We followed a slightly modified version of
the method described by \cite{GarciaBurillo2021} to fit 
the unresolved emission, which is assumed to provide an upper limit to the
synchrotron emission. Briefly, we started by fitting the extended
configuration $854\,\mu$m continuum map with a point source
with the observation beam size and orientation, and an elliptical
component. The estimated unresolved flux is $520\,\mu$Jy. We 
then subtracted the point-source fit from the
observations. The point-source subtracted image (Fig.~\ref{fig:torus},
top-right panel) shows no emission residuals
at the AGN position, that is, all the
observed emission at this angular resolution is unresolved. Next, we
generated a point source image corresponding to the
beam of the merged configuration data set, scaled it to the point
source flux from the previous step, and subtracted it from the
observations.

The $854\,\mu$m point source subtracted image shows significant
extended emission  at  the 3 to $5\times \sigma$ levels,
whereas at the AGN position there is no residual emission (bottom-right
panel of Fig.~\ref{fig:torus}). The measured fluxes
are $877\,\mu$Jy for the point + extended emission and
$364\,\mu$Jy after the point source subtraction. The
extended emission therefore accounts for approximately 42\% of the observed
flux within $r=32\,$pc and can be associated with cold dust emission in
the torus. It is likely that
the point source is oversubtracted since in the first step we assumed
that all the unresolved emission within the 11\,pc beam of the extended
configuration is produced by synchrotron emission.
Higher angular resolution observations
demonstrate that  at this far-IR wavelength, the synchrotron emission is confined to
regions of only a few parsecs in size
\cite[see][and 2023, in preparation]{GarciaBurillo2019,
  GarciaBurillo2021}.

We estimated the torus gas mass using the $854\,\mu$m extended emission flux and
assuming $T_{\rm dust} \simeq T_{\rm gas}=100\,$K,  a dust
emissivity value of $\kappa_{\rm 351GHz} = 0.0865\,{\rm m}^2\,{\rm
  kg}^{-1}$ \citep{Klaas2001}, and a gas-to-dust ratio of 100. We obtained $M_{\rm
  gas}^{\rm dust} \simeq 8.1 \times 10^5\,M_\odot$. For comparison, we integrated
 the CO(3-2) flux within the torus radius of 32\,pc and its derived
 inclination (see Table~\ref{tab:coldmoleculargas}),  using the 
same assumptions  as in  Sect.~\ref{subsec:outflowingring}. For this estimate, we took the
brightness temperature ratio measured in the NGC~ 1068 torus, $T_{\rm B}$CO(3-2)/$T_{\rm
  B}$CO(1-0)$=2.9$ \citep{GarciaBurillo2014}.  We derived a cold molecular
gas in the torus of $M_{\rm
  gas}^{\rm CO} \simeq 8.2 \times 10^5\,M_\odot$, in excellent 
agreement with the value estimated from the cold dust emission. 
It is also within the range of torus masses inferred in other Seyfert
galaxies \citep{GarciaBurillo2021}. We measured an  H$_2$ column
  density at the AGN position of $\log {\rm (N_{\rm H2}})=$22.34\,mol\,cm$^{-2}$,
  which agrees well the observed values in other GATOS Seyfert
  galaxies with similar AGN luminosities \citep[see Fig.~19, right,
  from][]{GarciaBurillo2021}.

\section{Discussion}\label{sec:discussion}

In Sect.~\ref{subsec:BarolodiskVRAD}, we showed that
the CO(3-2) minor axis p-v diagram (Fig.~\ref{fig:BaroloPVVRAD}) goes across significantly strong
emission from the ring. Under the commonly assumed hypothesis 
 that the bulk of molecular gas shares a coplanar geometry, the
 $v_{\rm rad}$ term is the only component of gas motions 
 that has a nonzero projection along the disk minor axis.
If the molecular gas is following circular orbits,  for the known orientation of the galaxy, the derived sign of the radial velocity
from the global $^{\rm 3D}$BAROLO  best fit to  the CO(3-2) data
cube implies the presence of a prevalent outward radial component
throughout the ring region and out to
$r\simeq700$~pc in the disk. In this section, we derive the properties
of this molecular outflow, compare them with those of the ionized
outflow, and  look for evidence of the impact of the outflow on
molecular gas in the
nuclear regions of NGC~7172.

\subsection{Properties of the cold molecular gas
  outflow}\label{subsec:outflowingring}

\cite{Stone2016}  also found tentative evidence for the
    presence of a molecular outflow in NGC~7172, using  $\sim 10\arcsec$ resolution
    Herschel/PACS observations of the OH  $119\,\mu$m absorption
    feature.
    They  measured an  
    approximate blueshifted velocity of $50\,{\rm km\,s}^{-1}$ of the OH
    feature, which the authors considered just at the limit of their outflow
    detection criterion. This value of the OH velocity is similar
    to the typical value of the CO(3-2) radial velocity fit with  $^{\rm 3D}$BAROLO in
    the molecular gas ring (see bottom panel of
    Fig.~\ref{fig:radialdistributions}).
    In the particular case of NGC~7172, the detection of the OH outflow
    was  probably aided by the fact
    that the outflow is observed close to our line of sight.

To compute the properties of the CO(3-2) molecular outflow, 
we assumed a
simple shell geometry
so that the mass outflow rate can be written as: 
 ${\rm d}M_{\rm gas}/{\rm d}t = M_{\rm gas} \times v_{\rm out}/R$
 \citep{Maiolino2012, Lutz2020}, where $M_{\rm
   gas}$ is the mass of molecular gas, $v_{\rm
   out}$ is the outflow velocity, and $R$ is the size of the 
 outflow region. For a coplanar geometry for the molecular outflow,
 the  outflow velocity is $v_{\rm out} = v_{\rm rad}$.
 To derive the molecular gas mass, we used equation~3 in
 \cite{Bolatto2013} with a CO-to-H$_2$ conversion factor of $X_{\rm
   CO} =2\times 10^{20}{\rm mol\,cm}^{-2} ({\rm K\,km\,s}^{-1})^{-1}$. This
 expression includes the helium correction.  To obtain the CO(1-0)
 intensity, we
 took a typical ratio of the brightness temperature in galaxy disks of
 $T_{\rm B}$CO(3-2)/$T_{\rm B}$CO(1-0)$=0.7$
 \citep[see][and references therein]{GarciaBurillo2021}. We
 integrated the CO(3-2) flux in the
ring (within $R=4\arcsec \simeq 720\,$pc, see
Table~\ref{tab:coldmoleculargas}), which provides a  molecular gas 
mass of $M_{\rm gas} =
6\times 10^8\,M_\odot$, and
taking an average value of the outflow velocity of $46\,{\rm
 km\,s}^{-1}$.   We inferred a mass outflow rate of  $40\pm 18\,M_\odot \,{\rm
    yr}^{-1}$, assuming that the main source of uncertainty in this
calculation comes from  the fitted
radial velocities.

\begin{figure}
     \includegraphics[width=9cm]{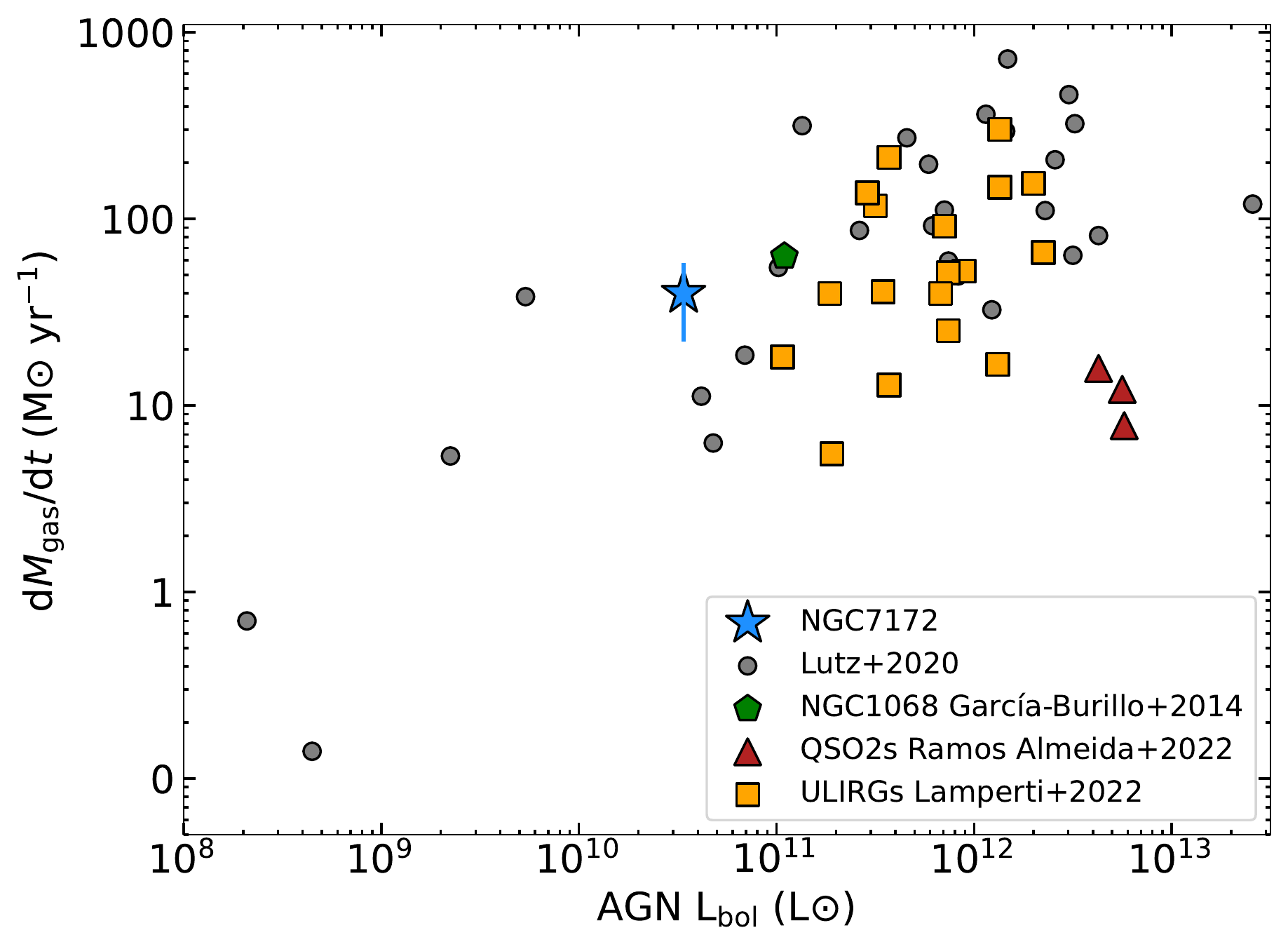}
\caption{Molecular gas mass outflow rate versus AGN bolometric
  luminosity of NGC~7172 compared with the compilation of infrared
  bright galaxies from
  \cite{Lutz2020} except for NGC~1068 which is from \cite{GarciaBurillo2014}, QSO2s from
  \cite{RamosAlmeida2022}, and local ULIRGs classified as AGNs from
  \cite{Lamperti2022}. For these comparison samples, we only plot
  galaxies with
  detections in both axes.}
              \label{fig:massoutflowratevsLbol}%
    \end{figure}

\subsection{Coupling between the AGN wind and the disk of the galaxy}

The molecular gas mass outflow rate of NGC~7172 derived in the
  previous section is large compared with the
{\it modest} ionized gas mass
outflow rate of 0.005\,M$_\odot$ yr$^{-1}$
\citep{Davies2020}. Interestingly, even though the Seyfert galaxies in their sample
have similar AGN luminosities, NGC~7172
has one of the lowest ionized mass outflow rates. 
This together with the observed scatter of the so-called AGN-wind scaling laws  might
 imply that  AGN luminosity is not the only
physical quantity explaining the different gas phases of  outflows in active
galaxies \citep{Fluetsch2019, Davies2020}.
On the other hand, our results indicate that NGC~7172 follows well the
trend between
$L_{\rm AGN}$ and molecular gas mass outflow rate found by
\cite{Lutz2020} for a compilation of local infrared bright galaxies,
 gray points in  Fig.~\ref{fig:massoutflowratevsLbol}. We note however
  that other works, such as \cite{RamosAlmeida2022}
   and \cite{Lamperti2022}, do not find tight correlations between these
  quantities for samples of type 2 quasars (QSO2s) and local
 ULIRGs classified as AGNs, red triangles and orange
  squares, respectively.

In simulations of AGN winds in disk galaxies, the outflows are predicted to
escape along the path of less resistance and thus are not able to
unbind completely a molecular disk \citep[see for instance,
][]{FaucherGiguere2012}.
However, outflows are predicted to be able to sweep up part of the material in
a galaxy disk, but at much lower velocities than the initial AGN wind,
which can be favored with a good geometrical coupling. An illustration
of the idea of the geometrical 
coupling can be found in Fig.~4 
of \cite{RamosAlmeida2022}.
 Observationally, the scarcity of strong
molecular outflows in several  Seyfert galaxies
\citep{DominguezFernandez2020}, QSO2s  with AGN-driven winds
\citep[][]{RamosAlmeida2022}, and local ULIRGs classified as AGNs \citep{Lamperti2022} has been
explained due to the lack of a  
geometrical coupling.

The derived  torus inclination of NGC~7172 is $\sim 61-67^\circ$
(see Sect.~\ref{sec:torus}),  while the molecular gas ring is seen almost edge-on
(see Sect.~\ref{sec:kinematics}). The inclination
(presumably perpendicular to the torus) and
wide opening angle of the
ionization cone of NGC~7172 \citep{Thomas2017} therefore allow for a relatively strong
geometrical coupling
between the AGN wind and the  galaxy interstellar medium (ISM), thus favoring the molecular
outflow in the ring. 
The  good spatial
correspondence between the CO(3-2), near-IR H$_2$ and continuum
$854\,\mu$m bright knot in  
the ring to the southwest of the AGN, and the edge of the ionization
cone traced by the VLT/SINFONI [Si\,{\sc vi}] emission
(Sect.~\ref{sec:morphology} and Figs.~\ref{fig:RINGALMACOcont} and
\ref{fig:ALMASINFONI}) add support to this scenario. At this particular
location of the NGC~7172 ring (also on the other side, to the
northeast of the AGN), it is likely that 
the impact with material in the ring caused a strong deceleration of
the AGN wind and subsequently in the outflow velocity of the
molecular gas, as seen in Fig.~\ref{fig:radialdistributions} (bottom panel). Indeed,
in 
\cite{Davies2020} sample,  NGC~7172 is among
the Seyfert galaxies with the lowest velocities of the ionized gas, thus
providing further evidence to the above scenario.

\subsection{Radio emission}\label{subsec:radioemission}

A radio jet, even with low power, running through a galaxy 
can interact with its ISM and entrain molecular gas in the galaxy
disk. Moreover, in simulations the geometrical coupling between  
radio jets and the ISM of 
Seyfert galaxies, including  its molecular
gas phase, has been found to be an important factor \citep{Mukherjee2018,
 Talbot2022}. Evidence of the interaction of radio jets with the
galaxy ISM has been observed in a few nearby Seyfert galaxies
\citep[see, e.g.,][]{GarciaBurillo2014, Morganti2015,
  PereiraSantaella2022, Peralta2023}.

In this section we discuss
the possibility that a low-power radio jet might be driving, at least in
part, the molecular outflow observed in NGC~7172.
The moderate angular resolution VLA 8.4\,GHz observations of
\cite{Morganti1999} revealed bright nuclear emission as well as 
extended emission along the major axis of the galaxy. This
extended emission likely corresponds 
to the molecular ring detected with our ALMA
observations. Subsequent VLA observations at subarcsecond resolution showed  again
nuclear
point-like emission and some faint
radio emission to the northeast and southwest of the AGN with an
approximate $2\arcsec$ extent \citep{Thean2000}. The orientation of
this faint extended radio emission, especially to the southwest, 
resembles that of the [Si\,{\sc vi}] emission \citep[see figure~8 of
][]{Smajic2012}.

We estimated a jet power of $P_{\rm jet}$$\sim$$10^{43}\,{\rm
  erg\,s}^{-1}$ for NGC~7172, using the relation given by 
\cite{Birzan2008} and
the 8.4\,GHz flux density quoted in \cite{Thean2000}, assuming a
power law $f_\nu \propto \nu^{-1}$ to derive the 1400\,MHz flux
density. The kinetic luminosity of the
molecular outflow can be computed as: $L_{\rm kin} = 1/2 \times
dM_{\rm gas}/dt \times v_{\rm out}^2$ \citep{GarciaBurillo2014}. We
obtained $L_{\rm kin} \sim   3 \times 10^{40}\,{\rm
  erg\,s}^{-1}$. If indeed, there is a radio jet in NGC~7172, it could
in principle provide sufficient power to drive the molecular outflow in this
galaxy. However, sensitive radio observations at higher angular
resolution are needed to confirm this.

\subsection{Feedback in the molecular gas ring of NGC~7172}\label{subsec:indexes}
In nearby active galaxies, feedback can manifest itself in various manners, including
suppressing star formation \citep{Gao2021},  triggering star formation
\citep{Cresci2015, Maiolino2017, Bessiere2022, Bellocchi2022}, as well as  redistributing the
molecular gas  in galaxy 
disks \citep{AlonsoHerrero2018, Shimizu2019, GarciaBernete2021,
  RamosAlmeida2022}. In a sample of GATOS Seyfert 
galaxies, \cite{GarciaBurillo2021} observed a clear imprint left by  the AGN wind on
the molecular gas at  torus and nuclear scales, with a decreased  molecular gas
surface density compared to that on circumnuclear scales ($r=200\,$pc) in
Seyfert galaxies with the highest AGN luminosities and/or Eddington ratios.

To evaluate the level of clearing of molecular gas in the torus and
nuclear regions of NGC~7172, we measured the cold molecular gas surface densities ($\Sigma$) on
different scales using the CO(3-2) fluxes reported in
Table~\ref{tab:coldmoleculargas}. 
We computed two cold molecular gas concentration indices
defined by \cite{GarciaBurillo2021}, namely,  Index-1=$\log \Sigma^{\rm
  r=50pc}_{\rm H2}/\Sigma^{\rm r=200pc}_{\rm 
  H2}=-0.16$ and  Index-II=$\log
\Sigma^{\rm torus}_{\rm H2}/\Sigma^{\rm r=200pc}_{\rm
  H2}=0.25$. The main difference between these two
  concentration indices in NGC~7172
is driven by the different inclinations used for the torus and the
$r=50\,$pc scales and the
different temperature brightness ratios (see Table~\ref{tab:coldmoleculargas}).
 The circumnuclear $r=200\,$pc scale does not
encompass the entire ring at all. Increasing the size of the circumnuclear
  region to include the ring gives a concentration index of   $\log \Sigma^{\rm
  r=50pc}_{\rm H2}/\Sigma^{\rm r=720pc}_{\rm 
  H2}=-0.6$.

The comparison of the two molecular gas concentration indices of NGC~7172
  with the sample of low-luminosity AGNs and GATOS Seyfert galaxies
  \citep[Fig.~18 of][]{GarciaBurillo2021} indicates
  that NGC~7172 follows well the observed trends for the two physical
  scales of the torus and its immediate surroundings. That is, there
  is a redistribution of the
  molecular gas in the torus and nuclear regions of NGC~7172 when compared to
  other less luminous Seyfert galaxies and AGNs. For this
  comparison, we used an  absorption-corrected $2-10\,$keV
luminosity of $\log L({\rm 2-10keV}) = 42.84$
  erg\,s$^{-1}$ \citep[][recalculated for the assumed distance in
  this work]{Ricci2017} and an Eddington ratio of $\sim 0.02$
  \citep{Vasudevan2010}.

  NGC~7172 shows molecular gas concentration
  indices similar to NGC~3227 and NGC~4388, which also show evidence
  for the presence of (circum)nuclear local outflows
  \citep{AlonsoHerrero2019,  DominguezFernandez2020,
    GarciaBurillo2021}. However, despite the clear
  evidence for the presence of a large-scale (out to $r\sim 4\arcsec
  \sim 720\,$pc from the AGN) outflowing ring in  NGC~7172 (see
  Sect.~\ref{subsec:outflowingring}), it 
  does not seem to present an {\it extreme} behavior in these
  relations. In other words, NGC~7172 does not show a markedly
  strong deficiency of cold molecular gas in its nuclear regions, as
  seen for instance in NGC~1068, even though both show comparable
  molecular gas outflow rates. One possible explanation is that the
  AGN wind is more
  collimated and faster in NGC~1068. Indeed, this galaxy shows an
  ionized gas outflow velocity of up to $1000\,{\rm km\,s}^{-1}$
  \citep{GarciaBurillo2019}, which is about 2.5 times faster than that
  in NGC~7172 \citep{Davies2020}. Furthermore,  the velocities of the
  molecular gas outflow in NGC~1068 are $100-150\,{\rm km \,s}^{-1}$
  \citep[see][]{GarciaBurillo2019},  thus about a factor of $2-3$
  faster than those
  in the outflow of NGC~7172. This likely resulted in a more efficient
  clearing of the molecular gas in the nuclear region of NGC~1068
  compared to NGC~7172.

 We finally explore the fate of the circumnuclear molecular gas
   in NGC~7172 since its
   mass outflow rate  appears to be at the high end of those
   observed in other AGNs of similar luminosity
   (Fig.~\ref{fig:massoutflowratevsLbol}). On the other hand, the velocity 
   of the molecular outflow of NGC~7172 is modest when compared with the galaxy
   circular velocity. In Appendix~\ref{sec:appendix2} we modeled the escape
   velocity in NGC~7172 using the $^{\rm 3D}$BAROLO rotation curve. At a radial distance of
   900\,pc, just outside of the outflow region, 
the escape velocity is $850\,{\rm km\,s}^{-1}$, whereas at smaller
radial distances this velocity is even higher (see Fig.~\ref{fig:vesc}).
Given that the molecular
   outflow is taking place mostly in the disk of the galaxy and its velocity is low with respect to the expected escape
   velocity, we can therefore conclude that in this galaxy most of the
 molecular gas will not leave the galaxy, at least during this period of AGN activity.

\section{Summary and conclusions}\label{sec:conclusions}
We presented new ALMA band 7 observations of the CO(3-2) transition
and associated $854\,\mu$m continuum, together with new VLT/SINFONI
observations of NGC~7172. This is a nearby
($d=37\,$Mpc) luminous
Seyfert galaxy in the GATOS sample with an AGN bolometric luminosity
of $1.3\times 10^{44}\,{\rm
  erg\,s}^{-1}$ \citep{Davies2015}.  The angular resolutions of the
ALMA observations ($0.06-0.3\arcsec$) allowed us to probe regions
on physical scales of $11-56\,$pc. 

The ALMA CO(3-2) intensity
maps show the presence of a highly inclined ($i\simeq 85-87^\circ$)
cold molecular gas ring with an
approximate radius of $3-4\arcsec \simeq 540-720\,$ pc, likely
associated with an ILR. There is also 
extended emission along the disk of the galaxy covering the FoV of the ALMA
observations along the east-west direction. This ring is
also detected at $854\,\mu$m and 
appears to trace the strong dust lane that  crosses the
(circum)nuclear region of NGC~7172. The cold molecular
gas mass of the ring is $\simeq 6.2\times 10^8\,M_\odot$.
The ALMA $854\,\mu$m continuum emission and the VLT/SINFONI H$_2$ at
  $2.12\,\mu$m and [Si\,{\sc vi}]$\lambda 1.96\,\mu$m line maps peak at
  the AGN position \citep[see][for a detailed discussion of previous
  SINFONI results]{Smajic2012}. The CO(3-2)
 map, on the other hand, shows the brightest emission in several
 clumps in the ring.  There are  also secondary H$_2$ and $854\,\mu$m
 peaks, 
 which additionally coincide with one of the bright CO(3-2) clumps in
 the ring. They all are located close to the edge of the 
 inner ionization cone delineated by  the [Si\,{\sc vi}] line emission.

The velocity fields of   [Si\,{\sc vi}], H$_2$, and CO(3-2) show
  clear evidence for the presence of noncircular motions. After
  subtraction of the stellar velocity field, the velocities of
    the ionized gas, traced with the  [Si\,{\sc
      vi}]$\lambda$$1.96\,\mu$m line, are blueshifted  a few hundred
 ${\rm km\,s}^{-1}$ in a region to the south of the AGN position. 
   We interpreted this as ionized gas
 outflowing outside the plane of the galaxy out to
 projected distances of $\simeq 200\,$pc. This emission likely
 corresponds to 
 the inner part of the larger ionized gas cone and outflow 
   detected in [O\,{\sc iii}]. The CO(3-2)
 p-v diagram along the kinematic minor 
  axis reveals the presence of noncircular motions of the cold
  molecular gas with observed
  velocities of up to $\sim 150\,{\rm km\,s} ^{-1}$. These are blueshifted
  to the   north of the AGN and redshifted to the south. 
Since the north side of the galaxy is the near
  side, these noncircular motions imply the presence of a molecular outflow in the
  plane of the galaxy. 

A $^{\rm  3D}$BAROLO model of a rotating disk with a radial 
  velocity component reproduces reasonably well the complicated
  CO(3-2) kinematics observed in NGC~7172.   We note that we also discussed an
  alternative scenario in which the noncircular motions could be the
  result of a nuclear bar (see Appendix~\ref{sec:appendix}). 
 However, the most plausible explanation for the observed  CO(3-2) 
kinematics indicates that the entire ring is not only rotating but also
  outflowing. The largest values of the fitted radial velocity take
  place close to the AGN with a gradual  deceleration,
  which is probably caused by the impact of the AGN wind, traced
  by the [Si\,{\sc   vi}] line, with the
  galaxy ISM.
  The integrated cold molecular gas mass outflow
  rate of the ring is $\sim 40\,M_\odot\,{\rm yr}^{-1}$, which is several
  orders of magnitude higher than that of the ionized phase
  \citep[$0.005\,M_\odot\,{\rm yr}^{-1}$, see][]{Davies2020}.
 The wide-angle ionization cone \citep[see 
Fig.~\ref{fig:ALMASINFONI}, and also ][]{Thomas2017} of NGC~7172 probably implies that the
AGN wind  is not highly collimated. As discussed by
\cite{MuellerSanchez2011}, this can result in  relatively {\it slow}
outflow velocities and moderate mass outflow rates of the ionized gas
phase, as seen in NGC~7172
\citep{Davies2020}. Despite this, the geometrical coupling between the wide-angle
ionization cone (perpendicular to $i_{\rm torus}$) and the disk of the
galaxy may have favored the interaction of the AGN wind with the surrounding ISM, thus
entraining molecular gas in the galaxy disk. 

The torus of NGC~7172 is detected as extended emission in the ALMA $854\,\mu$m
 map, after subtracting the unresolved emission. The latter likely has a
 synchrotron origin. The derived torus radius is 32\,pc and is seen at an
 inclination of $i_{\rm torus} \simeq 67^{\circ}$ (lower limit), based on the
 observed aspect ratio. The torus gas mass is
 $8\times 10^5\,M_\odot$, as measured  from the extended
 $854\,\mu$m emission as well as the CO(3-2) emission. These torus properties
 are similar to other Seyfert galaxies in the GATOS sample.  When
 compared to less luminous AGNs, NGC~7172
 presents decreased CO(3-2)  surface brightness on 
 torus and nuclear scales relative  to that on circumnuclear scales.  We
  interpreted this as evidence of redistribution of molecular gas in the
  torus and nuclear region of NGC~7172 due to AGN feedback, as
  observed in other luminous Seyfert galaxies.

  \begin{acknowledgements}
     We thank the referee for some insightful comments.
      AA-H and SG-B acknowledge support from grant
      PGC2018-094671-B-I00 funded by MCIN/AEI/ 10.13039/501100011033
      and by ERDF A way of making Europe. AAH also acknowledges
      support from PID2021-124665NB-I00 funded by the Spanish
  Ministry of Science and Innovation and the State Agency of Research
  MCIN/AEI/10.13039/501100011033 PID2021-124665NB-I00 and ERDF A way
  of making Europe.
  SGB also acknowledges support
      from the research project PID2019-106027GA-C44 of the Spanish
      Ministerio de Ciencia e Innovaci\'on.

      CRA acknowledges the projects ``Feeding and feedback in active galaxies,'' with reference
PID2019-106027GB-C42, funded by MICINN-AEI/10.13039/501100011033, ``Quantifying the impact of quasar feedback on galaxy evolution,'' with reference EUR2020-112266, funded by MICINN-AEI/10.13039/501100011033 and the European Union NextGenerationEU/PRTR, and from the Consejer\' ia de Econom\' ia, Conocimiento y Empleo del Gobierno de 
Canarias and the European Regional Development Fund (ERDF) under grant ``Quasar feedback and molecular gas reservoirs,'' with reference ProID2020010105, ACCISI/FEDER, UE.
CR acknowledges support from the Fondecyt Iniciacion grant 11190831
and ANID BASAL project FB210003. EB acknowledges the Mar\'ia Zambrano program of the Spanish Ministerio de Universidades funded by the Next
Generation European Union and is also partly supported by grant RTI2018-096188-B-I00 funded by MCIN/AEI/10.13039/501100011033.

This paper makes use of the following ALMA data: ADS/JAO.ALMA\#2019.1.00618.S. ALMA is a partnership of ESO (representing its member states), NSF (USA) and NINS (Japan), together with NRC (Canada), MOST and ASIAA (Taiwan), and KASI (Republic of Korea), in cooperation with the Republic of Chile. The Joint ALMA Observatory is operated by ESO, AUI/NRAO and NAOJ. 

Based on observations collected at the European Southern Observatory
under ESO programme 093.B-0057(B).

This research has made use of ESASky, developed by the ESAC Science
Data Centre (ESDC) team and maintained alongside other ESA science
mission's archives at ESA's European Space Astronomy Centre (ESAC,
Madrid, Spain).

This research has made use of the NASA/IPAC Extragalactic Database (NED),
which is operated by the Jet Propulsion Laboratory, California Institute of Technology,
under contract with the National Aeronautics and Space
Administration.

We acknowledge the usage of the HyperLeda database
(http://leda.univ-lyon1.fr).

This research made use of NumPy \citep{Harris2020}, Matplotlib \citep{Hunter2007} and Astropy \citep{Astropy2013, Astropy2018}.

\end{acknowledgements}

%
%

   \bibliographystyle{aa} 
   \bibliography{bibliography} 

   \begin{appendix}
  \section{Modelization of noncircular motions due to a central bar}
\label{sec:appendix}
This appendix describes a bar-ring model, with the nuclear bar
  being the source of noncircular motions in the center, as an
  alternative explanation for the observed CO(3-2) kinematics of the
  central regions of NGC~7172.
The central gas distribution mapped with ALMA has an 
extent of 3.2~kpc  or projected radius of 1.6~kpc (c.f., Fig.~\ref{fig:HSTALMAlarge}).
Its morphology reveals a central ring, with the characteristic enhancement
in the northeast and southwest parts of the ring, as shown in
Fig.~\ref{fig:RINGALMACOcont}. This indicates 
that the ring may be a pseudo-ring, made by the winding of the
two spiral arms, elongated along the bar, and usually observed
as the dusty leading edges of the bar in spiral galaxies. 
The projected radius of the ring is $\sim$ 300~pc, typical of
an ILR. This central molecular feature appears very elongated. 
It is likely however that the ring and barred structure
are intrinsically elongated, considering that the inclination of the galaxy is not
completely edge-on, but of the order of $\sim$ 60$^\circ$, as noted
by Hyperleda and the observed large-scale optical morphology 
(Fig.~\ref{fig:HSTALMAlarge}). We  then assumed this value of the inclination
in the following modelization.

\begin{figure}
\begin{center}
\includegraphics[width=0.6\textwidth,angle=0]{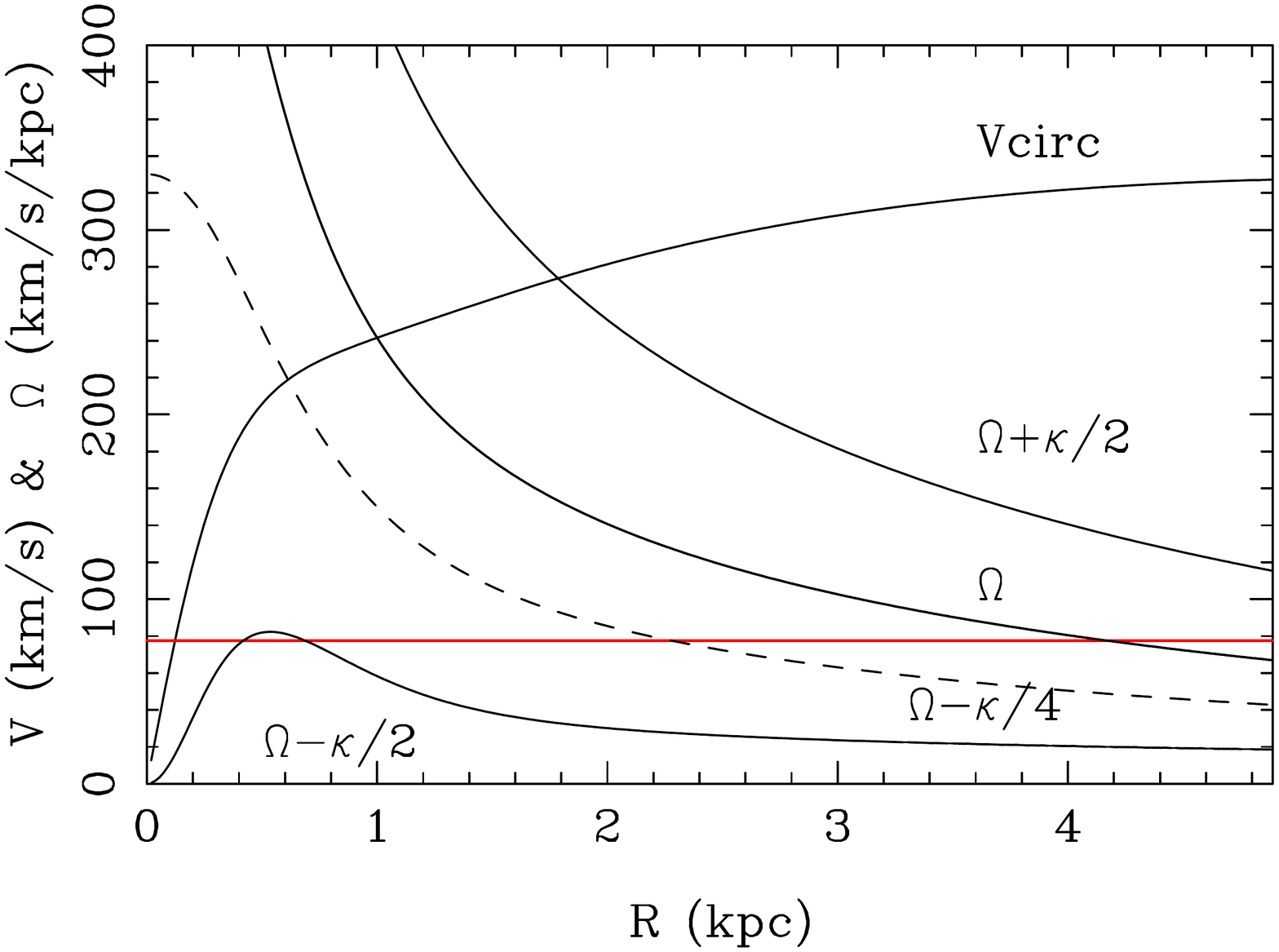}
\vskip+0.0cm
\caption{
      Circular velocity and corresponding 
	frequencies $\Omega$, $\Omega \pm \kappa/2$, $\Omega-\kappa/4$
        of the gas model for the center of NGC~7172. The pattern speed
	of the nuclear bar of the N-body simulation is marked in red.
}
\label{fig:vcur}
\end{center}
\end{figure}

\begin{figure}
    
\hspace{-2cm}
 \includegraphics[width=0.6\textwidth,angle=0]{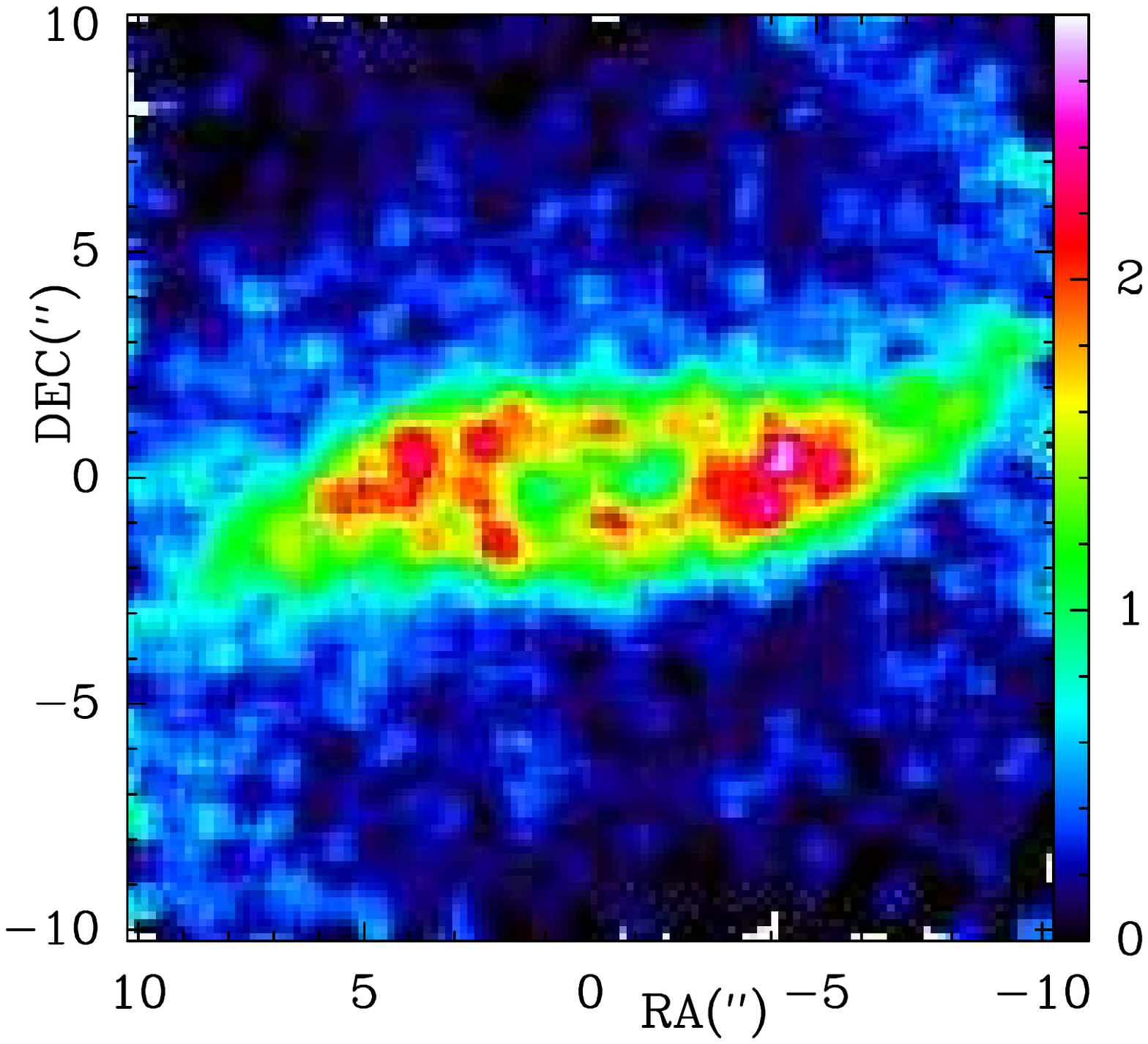}

 \vspace{-1cm}
\hspace{-2cm}
\includegraphics[width=0.6\textwidth,angle=0]{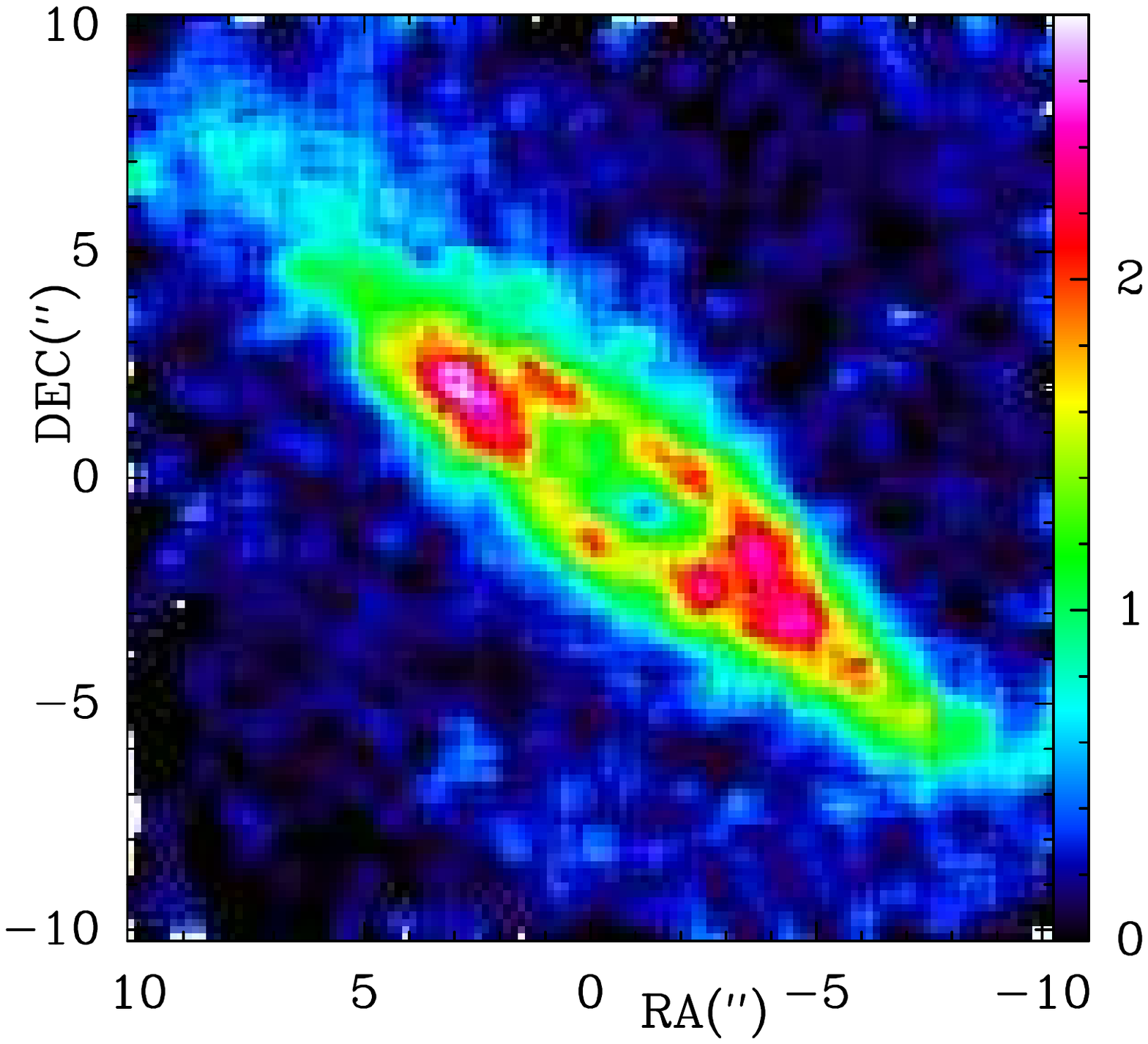}

\vspace{-1cm}
\hspace{-2cm}
\includegraphics[width=0.6\textwidth, angle=0]{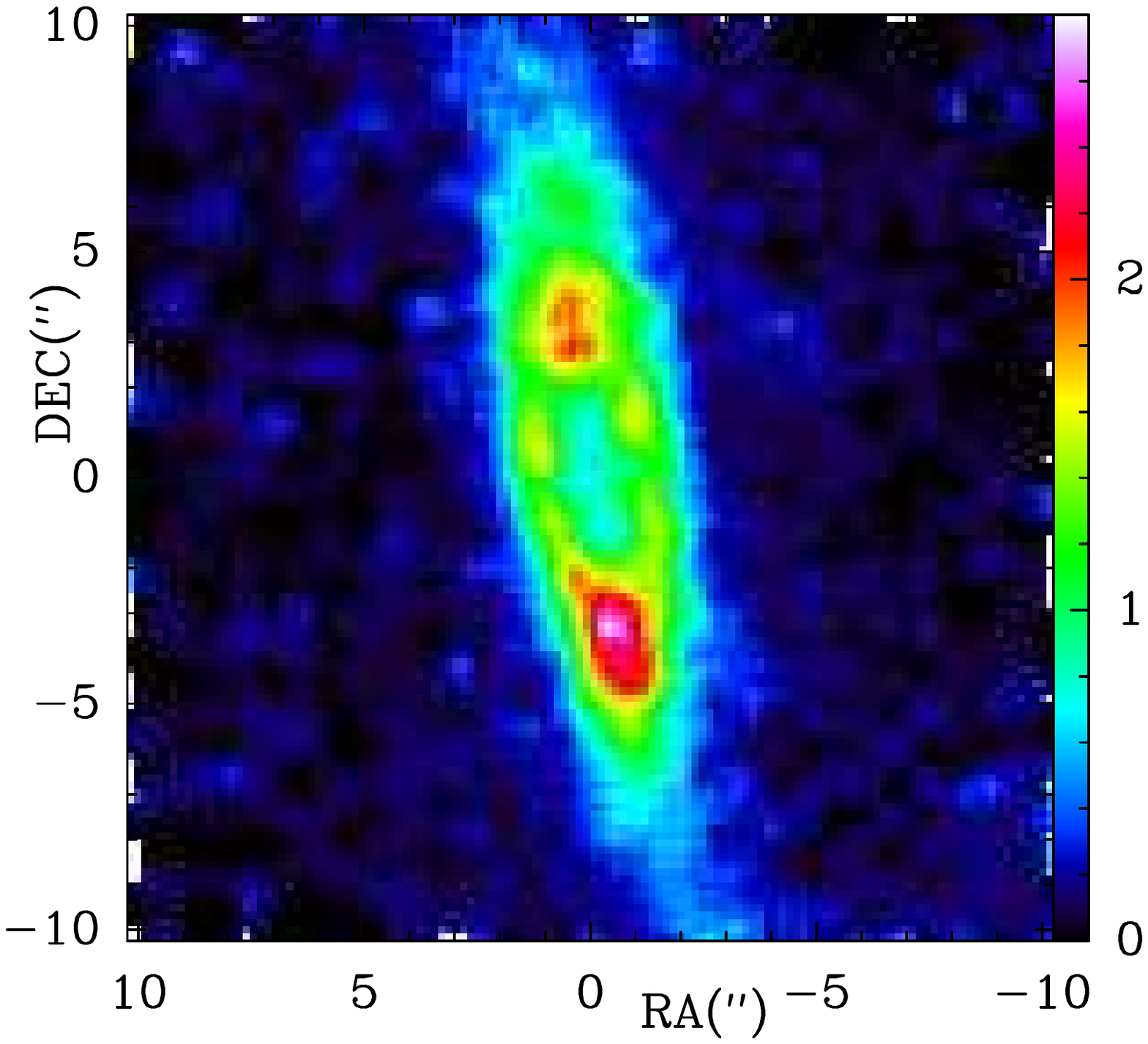}

\vspace{-0.75cm}
\caption{Three snapshots, separated by 33\,Myr, of the gas component
viewed face-on, for the NGC~7172 model.}
\label{fig:face}
\end{figure}

We used N-body simulations with gas, star formation, and feedback, as described in 
\cite{Combes2008}. The Poisson equation is solved with a 3D Particle-Mesh 
code based on FFT, and is fully self-consistent with
a live dark halo. The gas dissipation is represented
by sticky particles and a total of 240 000 particles is used. 
The star formation rate follows a Schmidt law, with exponent $n=1.4$,  a density
threshold, and an average gas consumption
time-scale of 5\,Gyr. The stellar feedback is modeled by  injecting
energy and gas velocity dispersion, when the stellar mass loss is 
distributed through gas on neighboring particles.
The potential of the galaxy is modeled with several components,
including a stellar bulge and disk, a dark matter halo, and a gas disk,
with the parameters listed in Table~\ref{tab:model}.
The shape of these components are Miyamoto-Nagai disks and Plummer
spheres, as described in \cite{Chilingarian2010}.
The computed circular velocity and angular frequencies,
 $\Omega$, $\Omega \pm \kappa/2$, $\Omega-\kappa/4$, are
 plotted in Fig.~\ref{fig:vcur}.

 \begin{figure*}

 \vspace{-1cm}
\hspace{-1cm}
\includegraphics[width=0.6\textwidth,angle=0]{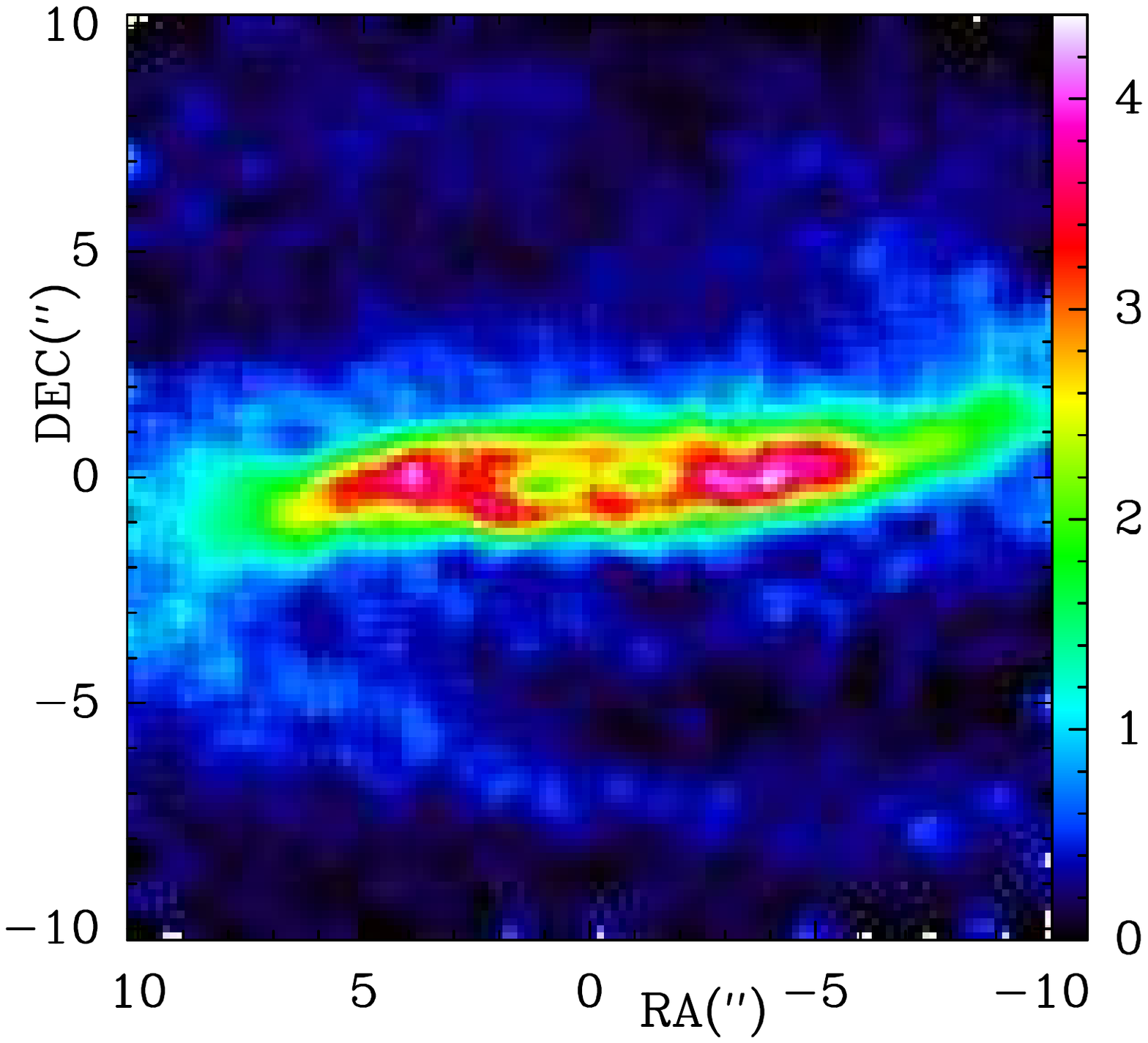}

\vspace{-7.5cm}
\hspace{9cm}
\includegraphics[width=0.55\textwidth,angle=0]{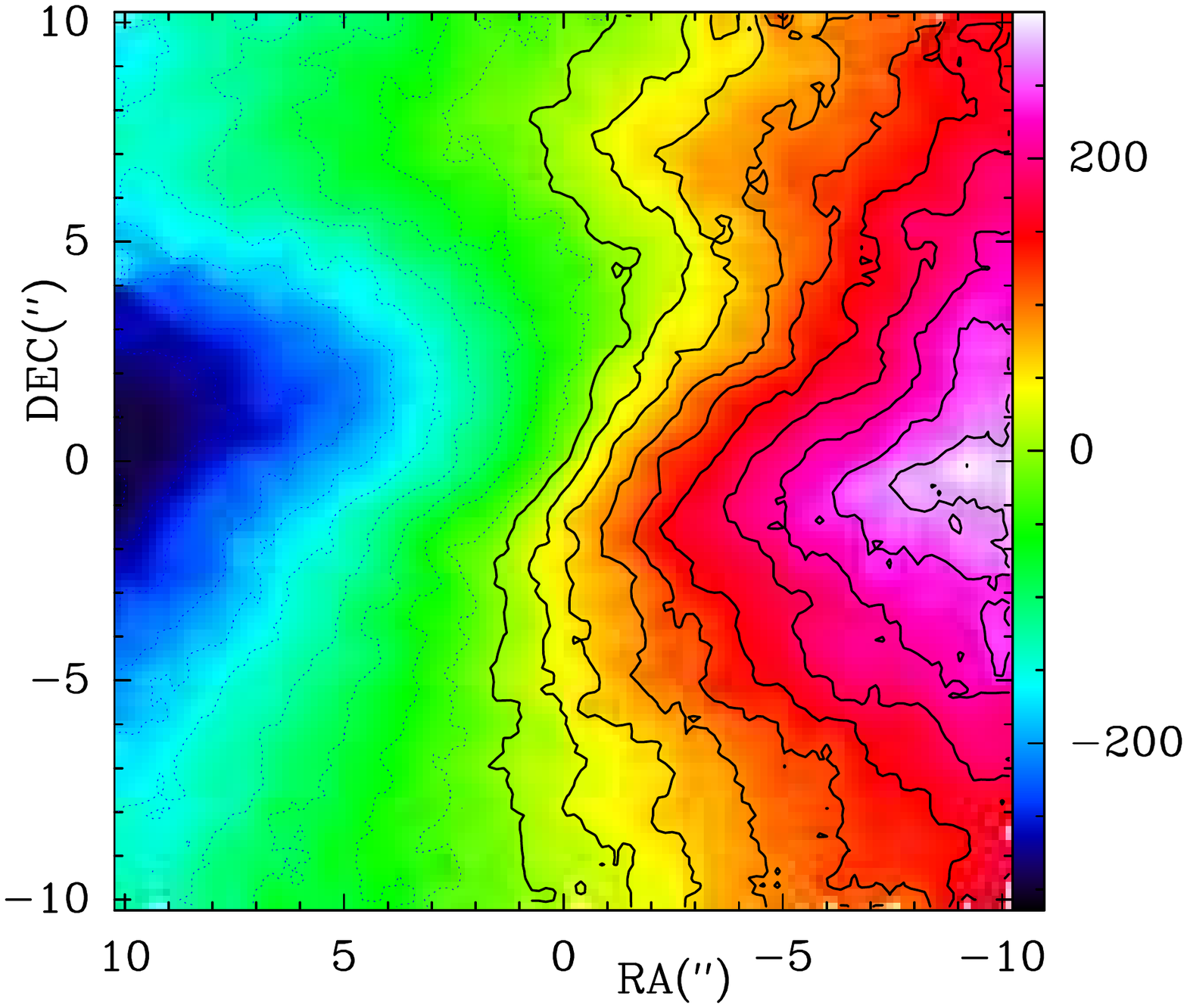}
\vspace{-0.2cm}
\caption{Moment maps of the gas model for the central
        regions of NGC~7172.
        {\it Left panel}. Moment 0. 
	The color scale is linear, in arbitrary units. {\it Right
          panel}. Isovelocities of the gas model for the central
        regions of NGC~7172.
	Negative velocity contours are traced by dotted lines.
	The spacing between isovelocities is 30\, km s$^{-1}$.
} 
\label{fig:mom0} 
\end{figure*} 

\begin{figure*}
    
\vspace{-5cm}
\hspace{1cm}
\includegraphics[width=0.45\textwidth,angle=0]{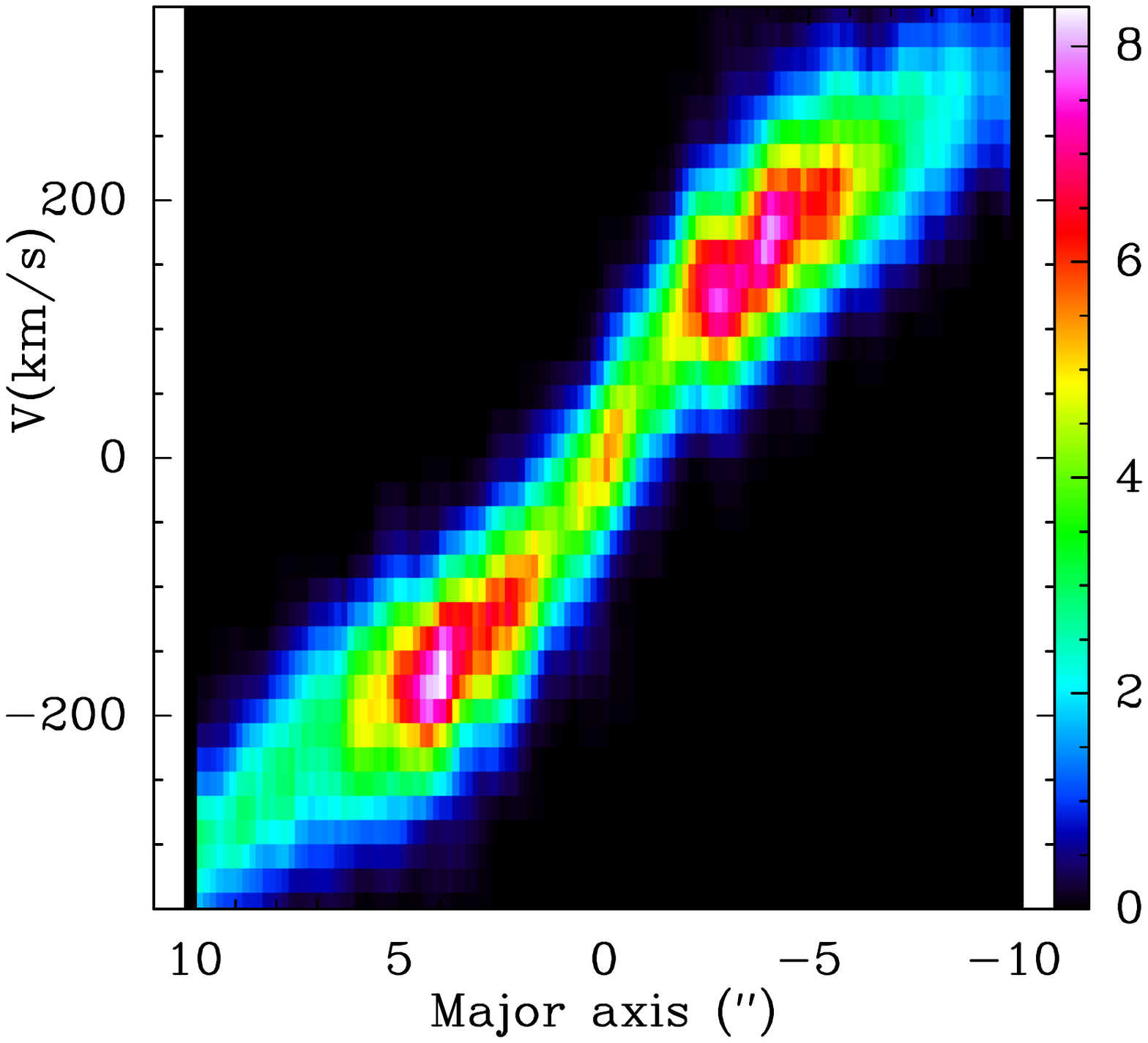}

\vspace{-12cm}
\hspace{9cm}
\includegraphics[width=0.45\textwidth,angle=0]{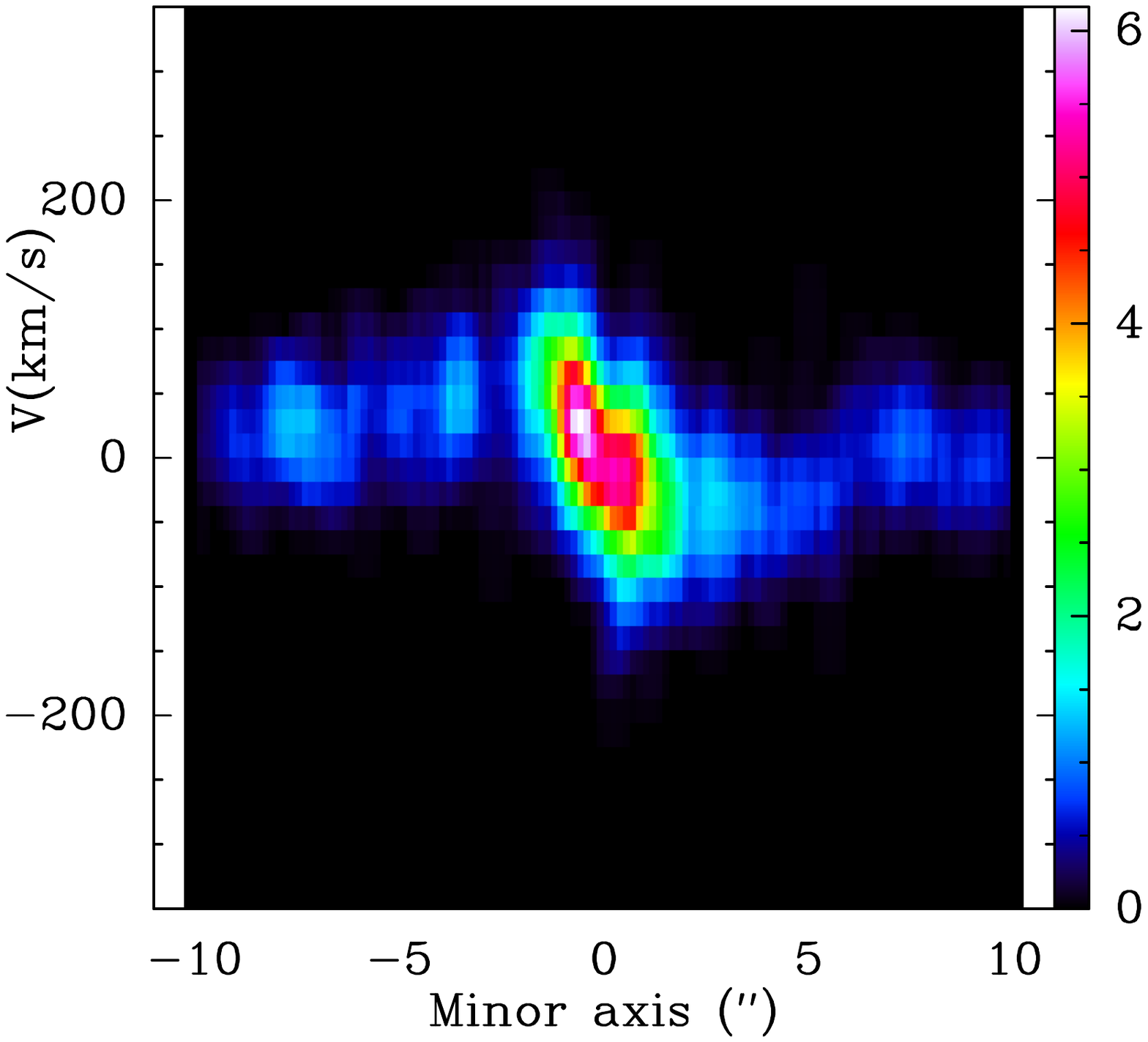}
\vskip+0.0cm
\caption{
       Computed p-v diagrams along the major axis
	(left) and minor axis (right)
	of the gas model for the central regions of NGC~7172.
	For the major axis, east is to the left, and west to the right. For the minor
	axis, south is to the left, and north to the right.
}
\label{fig:pvmai}
\end{figure*}

\begin{table}
\caption{Parameters adopted for the model.}
\vspace{-0.4cm}
\begin{center}
\begin{tabular}{ccccc}
\hline
\hline
Component              & Bulge & Disk  &  DM-Halo & Gas \\
\hline
Mass (10$^9\,$M$_\odot$)& 11.5  &  46   &  172     &  9  \\         
Radius (kpc)$^a$       & 0.44  &  2.22 &  5.33    &  2.6\\         
\hline
\end{tabular}
\tablefoot{ 
\tablefoottext{a}{Characteristic radius, see text for details.}  
        }
\label{tab:model}
\end{center}
\end{table}

In the simulation, after 300\,Myr a stellar bar has developed and  gas has accumulated
at the ILR, as shown in the three snapshots of Fig.~\ref{fig:face}.
The pattern speed of the bar $\Omega_p$ is $\sim$ 80\,km s$^{-1}$
kpc$^{-1}$, allowing barely for one ILR.

The gas component at the first snapshot of Fig.~\ref{fig:face} (top panel)
is then projected on the sky, with an inclination of   60$^\circ$,
and a position angle providing the best fit to the observations. 
The data are projected into a cube, with pixel sizes of $0.16\arcsec$
and 18\,km s$^{-1}$. We smoothed the cube to a beam size comparable to the compact ALMA
configuration of our band 7 observations.
We show the moment 0 of the cube  in Fig.~\ref{fig:mom0} (left panel)
and the isovelocities in Fig.~\ref{fig:mom0} (right panel).

The overall morphology of the molecular ring is reproduced in this simulation,
although the model does not intend to be a fit for this peculiar galaxy.
The simulations presented here are to demonstrate that a bar and ILR ring model, with its
noncircular motions, is able to reproduce the main observed features 
of the molecular gas in NGC~7172.
 The velocity field, with its characteristic S-shape frequently observed
 in barred galaxies, indicates noncircular motions corresponding to
 an apparent outflow of the molecular gas. Indeed, the near-side of the galaxy is 
 to the north, and far side to the south. This is deduced from the dust morphology,
 and corresponds also to a trailing spiral pattern.
Along the minor axis, the velocity is
 negative with respect to the systemic velocity, 
the gas is moving toward the observer to  the north, and the reverse to the south
This apparent gas outflow is also illustrated in the p-v diagrams of 
Fig.~\ref{fig:pvmai}. The major axis is at PA=90$^\circ$, and the orientation
of the cuts is indicated in the caption. We notice that this apparent outflow
is particularly deceiving, since the gas in the simulation is in fact inflowing,
due to the gravity torques of the stellar bar.

\section{Escape velocity}\label{sec:appendix2}
\begin{figure}

 \vspace{-2cm}
\hspace{-1cm}
\includegraphics[width=1.1\textwidth,angle=0]{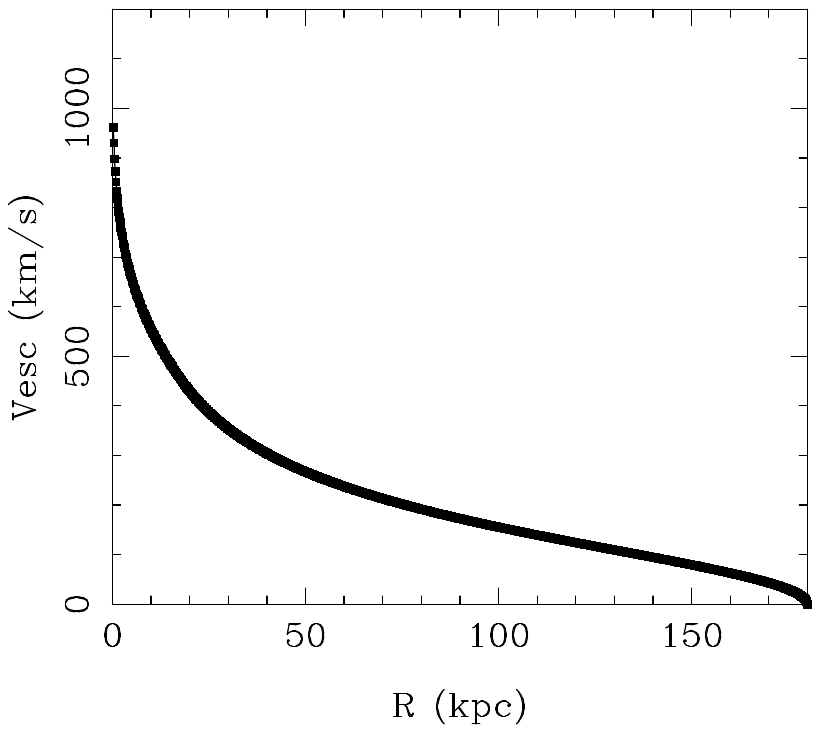}
\vspace{-18cm}
\caption{Modeled escape velocity for NGC~7172.}

\label{fig:vesc} 
\end{figure}

 In this appendix we estimate the escape velocity ($V_{\rm esc}$) as a function of the
radial distance for NGC~7172. For a particle at radius $R$, located in
a gravitational potential   $\Phi(R)$, the definition of the escape
velocity is when the kinetic energy per unit mass is
$V^2/2 = - \Phi(R)$.  It is assumed that  $\Phi(R)$ reaches zero in the outskirt
of the potential, that is, the virial radius of the galaxy. From the
rotation curve $V_{\rm rot}$
of the galaxy, estimated from the observed p-v diagram of
Fig.~\ref{fig:BaroloPVVRAD} modeled with $^{\rm 3D}$BAROLO, we can estimate
the potential, assuming in a first approximation spherical symmetry.
We can thus write   $V_{\rm rot}^2/r =  {\rm d}\Phi(r)/{\rm d}r$, and compute:

$$\Phi(R) = \Phi_0 + \int_{0}^{R} V_{\rm rot}^2(r) dr/r$$

\noindent where $\Phi_0$ is then computed throughout $\Phi(R_{\rm vir})$ = 0.
We have adopted a simple rotation curve, taking a linearly increasing
$V_{\rm rot}$ from 0 to $300\,{\rm km\,s}^{-1}$ up to
2.5\arcsec=0.45\,kpc, and then a constant 
$V_{\rm rot}$ =$300\,{\rm km\,s}^{-1}$ all the way to $r$=100\arcsec=18\,kpc. After this optical
radius, we assume that the dark matter halo is dominating, with
a  Navarro-French-White profile, where the dark matter density is
varying as $\rho(r) \propto r^{-3}$, and the potential becomes $\propto {\rm log}(r)/r$
from 18\,kpc  to $R_{\rm vir}$ =180\,kpc (virial radius).
At each radius $V_{\rm esc}^2 (R)= 2 |\Phi(R)|$. Figure~\ref{fig:vesc}
shows the derived $V_{\rm esc}$ as a function of the radial distance. 
At  0.9\,kpc, $V_{\rm esc}$ is already $850\,{\rm km\,s}^{-1}$.

\end{appendix}

 \end{document}